\RequirePackage{fix-cm}
\documentclass[a4paper,fleqn]{cas-sc}
\usepackage{graphicx}
\usepackage{float}
\usepackage{threeparttable}
\usepackage{url}
\usepackage{nccmath}
\usepackage{diagbox}
\usepackage{ulem}
\usepackage{caption}
\usepackage{algorithm}%
\usepackage{algpseudocode}%
\usepackage{subcaption}
\usepackage[authoryear]{natbib}
\def\tsc#1{\csdef{#1}{\textsc{\lowercase{#1}}\xspace}}
\tsc{WGM}
\tsc{QE}
\tsc{EP}
\tsc{PMS}
\tsc{BEC}
\tsc{DE}
\usepackage{appendix}
\hypersetup{
    colorlinks=true,
    citecolor={RoyalBlue},
    linkcolor={RoyalBlue},
    urlcolor={RoyalBlue}
}
\usepackage{etoolbox}
\ExplSyntaxOn
\cs_if_exist:NF \msg_term:n { \cs_new_protected:Npn \msg_term:n #1 { \iow_term:n {#1} } }
\ExplSyntaxOff

\begin{document}
\begin{sloppypar}
\let\WriteBookmarks\relax
\def\floatpagepagefraction{1}
\def\textpagefraction{.001}
\shorttitle{}
\shortauthors{Hao Lyu et~al.}

\title [mode = title]{Understanding traffic flow evolution in mixed autonomy traffic through trajectory-informed, traffic-state-conditioned behavioral parameterization}
\author[1,2]{Hao Lyu}[orcid=0000-0002-1664-8050]
\ead{lyu_hao@seu.edu.cn}

\address[1]{School of Transportation, Southeast University, Nanjing, 211189, China}
\address[2]{Mobility and Industrial Management~(MIM) Division, KU Leuven, Leuven, 3001, Belgium}

\author[1]{Yanyong Guo}
\cormark[1]
\ead{guoyanyong@seu.edu.cn}

\author[2]{Claudio Roncoli}
\ead{claudio.roncoli@kuleuven.be}

\cortext[cor1]{Corresponding author}



\begin{abstract}
Mixed traffic flow analysis depends on how heterogeneous car-following behavior is parameterized, yet many studies use fixed or transferred parameters that are not aligned with the local traffic states being analyzed. This study develops a trajectory-informed and traffic-state-conditioned behavioral parameterization. Car-following interaction segments from large-scale trajectory data of human-driven vehicles (HVs) and automated vehicles (AVs) are calibrated with the Intelligent Driver Model (IDM), and the calibrated parameter samples are linked to local traffic-state descriptors and follower--leader interaction types to learn a conditional parameter representation. Validation shows that the calibrated parameters reproduce short-horizon car-following behavior and that the generated parameters reproduce the observed distributions of trajectory-level indicators and support one-shot long-horizon simulation. System-level experiments with the generated parameters, conducted under controlled AV penetration rates, clustering intensities, and speed conditions, reveal two mechanisms through which AV car-following behavior shapes mixed traffic. The first is the equilibrium headway: AV-follower interactions have longer headways than HV-follower interactions at low speed but shorter headways at medium and high speeds, so higher AV penetration reduces equilibrium flow at low speed and increases it at medium and high speeds. In ring-road simulations, this reversal raises the capacity point at medium and high speeds but leaves pure AV traffic with the lowest congested branch. The second is the disturbance response: AV-follower interactions are more prone to low-frequency amplification in most examined conditions, and their repetition under positive clustering accumulates into stronger head-to-tail amplification and a sharper post-critical flow reduction with persistent upstream-moving low-speed waves. The effect of AV penetration is therefore conditional on speed and composition and involves a trade-off between flow efficiency and stability.
\end{abstract}

\maketitle
\section{Introduction}

As automated driving technologies continue to advance, road traffic is expected to experience a prolonged transitional phase in which human-driven vehicles (HVs) and automated vehicles (AVs) coexist~\citep{yu2021automated, 11077689}. In this mixed traffic environment, system-level outcomes are shaped not only by the behavior of individual AVs, but also by how heterogeneous interactions involving HVs and AVs are organized along the traffic flow~\citep{zhong2020influence}. Mixed traffic flow analysis has therefore become important for examining how automation affects capacity, stability, disturbance propagation, and traffic-flow evolution. Since real-world deployment conditions cannot be exhaustively observed yet, these questions can be studied through analytical models and microscopic simulations under controlled traffic compositions, scenario settings, and behavioral assumptions~\citep{yao2023analysis}.

A central difficulty in such analyses is how microscopic car-following behavior is parameterized. Many mixed traffic studies specify HV and AV behavior using fixed, assumed, transferred, or weakly matched parameter settings~\citep{qin2023stability, yao2023analysis}. These practices are convenient, but they can obscure the behavioral source of system-level differences. Car-following behavior varies across follower--leader interaction types, such as HV--HV, HV--AV, AV--HV, and AV--AV, and may also vary within the same interaction type under different local traffic states~\citep{li2023large, wu2025modeling}. If these differences are represented by parameters obtained from different datasets, road environments, calibration settings, or operating contexts, the resulting comparison between HVs and AVs may partly reflect unmatched behavioral inputs rather than vehicle-type effects alone. This concern is reinforced by evidence that car-following calibration results can be sensitive to data source, calibration setting, and estimation procedure~\citep{punzo2021calibration}. Divergent behavioral assumptions have also led to different conclusions regarding traffic stability, throughput, and bottleneck dynamics in mixed traffic~\citep{shang2021impacts, luo2024stabilizing}. Behavioral parameterization is therefore a methodological component of mixed traffic analysis, not a secondary simulation detail~\citep{punzo2020two}.

Recent high-resolution trajectory datasets provide an opportunity to ground this parameterization in empirical evidence. AV-oriented open datasets and large-scale trajectory collections make it increasingly feasible to examine mixed HV--AV interactions using empirical evidence under more specific traffic states and interaction conditions~\citep{li2020trajectory, ivanovic2023trajdata, zhang2024calibrating}. At the same time, empirical data alone cannot cover the broad experimental space required for mechanism-oriented mixed traffic analysis. Questions involving wide ranges of AV penetration rates, clustering intensities, speed conditions, boundary settings, and density conditions still require controlled simulation experiments~\citep{li2025influence}. A feasible direction is therefore to construct behaviorally grounded model inputs from trajectory evidence and then deploy these inputs in comparable system-level simulations.

The remaining methodological gap is the lack of a unified way to derive behavioral parameter samples that are comparable across interaction types, conditioned on the local traffic states being analyzed, and directly usable in mixed traffic simulation. Existing studies have made important progress in identifying heterogeneity from observed trajectories~\citep{yu2026mcvae, kong2026active}, but less attention has been paid to translating this heterogeneity into conditional parameter representations that can be supplied consistently under prescribed traffic states and follower--leader interaction types. Motivated by this gap, this study develops a trajectory-informed and traffic-state-conditioned behavioral parameterization framework for mixed traffic, where observed interaction segments are used to calibrate the same physical car-following model across all four interaction types, yielding a parameter sample for each segment. These samples are linked to short-term traffic-state descriptors and interaction types, and a conditional representation is learned to generate behavioral inputs under specified local traffic states. The generated parameters are then deployed in comparable mixed traffic analyses to examine how AV penetration rate, clustering intensity, speed condition, and boundary or density setting jointly affect system-level evolution.

The main contributions of this study are summarized as follows:
\begin{enumerate}
    \item We develop a behavioral parameterization method for heterogeneous car-following interactions in mixed traffic. The proposed method transforms observed interaction segments into calibrated physical-model parameter samples, links them with local traffic-state descriptors and follower--leader interaction types, and learns a conditional representation for generating behavioral inputs.

    \item We incorporate the proposed behavioral parameterization into a comparable mixed traffic analysis workflow. Vehicle sequences are constructed by jointly considering AV penetration rate and clustering intensity, and follower-specific behavioral parameters are deployed through conditional sampling based on the corresponding initial local traffic states and interaction types.

    \item We identify two mechanisms, equilibrium headway and disturbance response, through which AV car-following behavior shapes mixed traffic, and show how AV penetration rate and clustering intensity, through these mechanisms, affect equilibrium flow, capacity, stability, and ring-road flow evolution under different speed conditions.
\end{enumerate}

The remainder of this paper is organized as follows. Section~\ref{s2} reviews related work on car-following modeling, heterogeneous parameterization, and system-level mixed traffic evaluation. Section~\ref{s3} defines the research scope and presents the overall framework. Section~\ref{s4} introduces the trajectory-informed and traffic-state-conditioned behavioral parameterization method. Section~\ref{s5} describes mixed traffic state construction, parameter deployment, experiment design, and evaluation metrics. Section~\ref{s6} validates the behavioral parameterization using calibrated parameter samples and generated parameters. Section~\ref{s7} presents the system-level results on equilibrium headway, disturbance response, and ring-road flow evolution. Section~\ref{s8} discusses the underlying mechanisms and their implications, and Section~\ref{s9} concludes the study and outlines future research directions.

\section{Related work\label{s2}}

\subsection{Microscopic car-following modeling for mixed traffic analysis}

Microscopic car-following modeling provides a fundamental basis for mixed traffic analysis, because many traffic-level phenomena originate from local longitudinal interactions among adjacent vehicles~\citep{qin2023stability}. Under HV and AV coexistence, changes in follower--leader interaction mechanisms can be transmitted to capacity, analytical stability, disturbance propagation, and traffic-flow evolution~\citep{yao2023analysis, 11077689}. Car-following models are therefore used not only to reproduce vehicle trajectories, but also to support mechanism-oriented analysis of mixed traffic systems.

Recent car-following research has introduced data-driven models~\citep{zhou2022transfer}, physics-informed or hybrid frameworks~\citep{lyu2025knowledge}, and control- or learning-based approaches such as model predictive control and reinforcement learning~\citep{hart2024towards, lyu2025mitigating}. These methods have shown strong potential in trajectory prediction, controller design, and behavior representation~\citep{zhang2024car}. For mixed traffic studies that require scalable microscopic simulation and interpretable traffic-flow analysis, physics-based car-following models remain widely used because their parameters and stability properties can be directly linked to capacity, disturbance propagation, and traffic-wave mechanisms~\citep{guan2025jam, li2025influence}.

Within this physics-based modeling line, mixed traffic studies have mainly followed two directions. One extends or recalibrates classical car-following models, including the Optimal Velocity Model, the Full Velocity Difference Model, the Gipps model, Newell-type models, and the Intelligent Driver Model (IDM). The other adopts dedicated adaptive cruise control (ACC) or cooperative adaptive cruise control (CACC) formulations for automated following behavior~\citep{milanes2014modeling}, sometimes combined with classical car-following models in unified mixed traffic frameworks~\citep{zhang2016motif}. Among these formulations, IDM is frequently used because its compact nonlinear structure can represent free flow, congestion, stop-and-go oscillations, and stability-related phenomena~\citep{hu2023autonomous, li2023automated, zhou2025twenty}.

These studies show that physical car-following models provide an interpretable and tractable basis for mixed traffic analysis. At the same time, behavioral heterogeneity within a common model is introduced primarily through parameterization~\citep{punzo2020two}. If HV and AV behaviors are represented by fixed, assumed, borrowed, or sparsely calibrated parameter sets~\citep{yao2023analysis}, the resulting simulations may be too coarse to capture interaction-specific behavior across follower--leader types and local traffic states. This motivates more empirically grounded behavioral parameterization for existing car-following models.

\subsection{Calibration and conditional generation of heterogeneous car-following parameters}

The parameterization problem has long been recognized in car-following research~\citep{ossen2011heterogeneity}. Early mixed traffic studies often used assumed, borrowed, or roughly estimated parameter settings to represent AV capabilities and HV responses. Under limited empirical evidence, this provided a practical basis for scenario exploration and mechanism-oriented analysis~\citep{ntousakis2015acc,talebpour2016influence, ghiasi2017mixed, papamichail2019review, li2022trade}. However, fixed parameter settings can only represent heterogeneity coarsely, making it difficult to separate vehicle-type effects, interaction-specific effects, and traffic-state effects in a consistent way.

With the growth of high-resolution trajectory data, car-following parameterization has increasingly shifted from assumed parameters to empirically calibrated parameters~\citep{kesting2008calibrating}. Parameters have been estimated from observed trajectory segments through optimization-based~\citep{chen2010calibration} and statistical approaches~\citep{hoogendoorn2010calibration}, with the common goal of improving trajectory fidelity while preserving physical interpretability. However, calibration alone does not fully resolve the comparability issue~\citep{punzo2012can}. Parameter estimates can depend on data source, traffic environment, objective function, and calibration procedure, so calibrated results from different studies are not necessarily comparable even when the same car-following model is used~\citep{punzo2016speed, punzo2021calibration}.

More recent work has moved toward controlled and heterogeneity-aware calibration settings. Existing studies have examined differences between HV and AV car-following behavior under more consistent empirical conditions~\citep{hu2023autonomous}, follower--leader interaction types~\citep{durrani2016calibrating, reddy2026calibration}, state-dependent behavior~\citep{yao2025novel}, driver styles~\citep{wu2025modeling}, random coefficients~\citep{zhong2024understanding}, and time-varying parameters~\citep{yu2026mcvae}. These studies provide important evidence that mixed traffic behavior is heterogeneous and condition dependent. At the same time, their outputs are often tied to finite trajectory samples, discrete groupings, or specific experimental settings, which limits their direct use as scalable behavioral inputs for broad mixed traffic simulations.

This limitation becomes more important when mixed traffic analysis spans many AV penetration rates, clustering intensities, interaction compositions, and traffic conditions~\citep{yao2023analysis}. Repeated scenario-specific calibration is impractical under such requirements. This motivates parameter generation, where generation refers not to generic trajectory prediction~\citep{zhang2022generative}, but to supplying heterogeneous car-following parameters for simulation initialization under specified local traffic states and follower--leader interaction types. Related studies have begun to combine physics-based behavioral models with generative or inferential architectures, but many still rely on complete historical trajectories, long sequential inputs, or relatively heavy model structures~\citep{kang2023trajectory, lyu2025presimnet}. The remaining challenge is to translate empirically observed heterogeneity into a traffic-state-conditioned conditional parameter representation that can provide comparable behavioral inputs for mixed traffic simulation.

\subsection{Mixed traffic composition, boundary settings, and system-level evaluation}

Beyond microscopic behavior representation, mixed traffic analysis also depends on how heterogeneous vehicles are organized into system-level experiments~\citep{zhao2025micro}. This analytical layer is typically defined through traffic composition, roadway or boundary setting, and system-level evaluation metrics~\citep{yao2023analysis}. Together, these choices determine how mixed traffic conditions are represented and how their macroscopic consequences are interpreted.

Mixed traffic composition is no longer described only by AV penetration rate. Recent analyses also consider platoon size~\citep{yao2023analysis}, clustering intensity~\citep{ghiasi2017mixed, li2026unveiling}, spacing-mode probabilities~\citep{sun2024fundamental}, and heterogeneous follower--leader combinations~\citep{kumar2026macroscopic}. These variables characterize not only the expected proportion of AVs, but also the ordering and continuity of different vehicle types along the traffic flow. For example, \citet{ghiasi2017mixed} showed that different clustering intensities can produce different capacity outcomes at the same penetration rate. Mixed traffic compositions are also evaluated under different roadway or boundary settings, including ring-road, straight-road, bottleneck, and merging scenarios~\citep{yao2023analysis, qin2023stability, zhao2025micro}. Therefore, system-level conclusions depend on both composition and scenario representation.

System-level evaluation has also expanded from capacity and macroscopic flow characteristics to a broader set of outcomes. Early studies emphasized throughput and fundamental diagrams~\citep{talebpour2016influence, zhou2020modeling}. Later work incorporated analytical stability, disturbance amplification, traffic-wave evolution, and jam formation~\citep{qin2023stability, yao2023analysis, zhao2025micro}, and some studies also considered safety and environmental indicators~\citep{jiang2025safety, dong2025analyzing}, showing that these outcomes may respond differently to the same mixed traffic composition. Boundary setting is also influential: for example, \citet{cummins2021simulating} showed that a wave-dissipation strategy can perform differently under ring-road and straight-road scenarios. This indicates that traffic composition, boundary setting, and evaluation dimension must be considered jointly.

As this analytical framework becomes more refined, the role of behavioral inputs becomes more consequential. \citet{montanino2021homogeneous} showed that homogeneous treatment can overestimate string instability in heterogeneous traffic, while validation evidence indicates that homogeneous or average-driver settings can generate traffic patterns that deviate from observed behavior~\citep{punzo2020two}. Empirical studies also show that car-following behavior is conditional on traffic state; for instance, \citet{hu2023autonomous} found that time headway changes with speed, which can affect capacity-related analysis in mixed traffic~\citep{li2026unveiling}. Similarly, \citet{qin2023stability} emphasized that empirical mixed-flow stability analysis should rely on car-following models calibrated from real experimental data and should be interpreted under the corresponding experimental setting.

Overall, existing research has established the importance of physical car-following models, heterogeneous parameter calibration, and system-level mixed traffic evaluation, but these strands are often treated separately. Model-based system analyses may use behavioral parameters that are not aligned with the local traffic states under evaluation, while calibration-oriented studies may identify heterogeneity without providing a direct parameter deployment interface for broad mixed traffic simulations. This leaves a methodological gap between empirical trajectory evidence and comparable system-level analysis. The present study addresses this gap by developing a trajectory-informed and traffic-state-conditioned behavioral parameterization framework and deploying the resulting generated parameters under controlled variations in AV penetration rate, clustering intensity, speed condition, and boundary or density setting.

\section{Problem definition and overall framework\label{s3}}

\subsection{Research scope and problem definition}

This study focuses on the system-level evolution of mixed traffic flow composed of HVs and AVs. The analysis is confined to the longitudinal component of mixed traffic dynamics, where continuous car-following interactions between adjacent vehicles transmit local behavioral differences to traffic-flow outcomes. Lateral maneuvers and lane-changing decisions are beyond the scope of this study.

Within this scope, mixed traffic is organized through four follower--leader interaction types, namely HV--HV, HV--AV, AV--HV, and AV--AV. The local longitudinal interactions associated with these interaction types are represented through car-following parameterizations learned from mixed traffic trajectory data. These parameterizations provide the behavioral basis for incorporating heterogeneous interactions into mixed traffic analysis.

\subsection{Proposed framework}

\begin{figure}[pos=t]
    \centering
    \includegraphics[width=\textwidth]{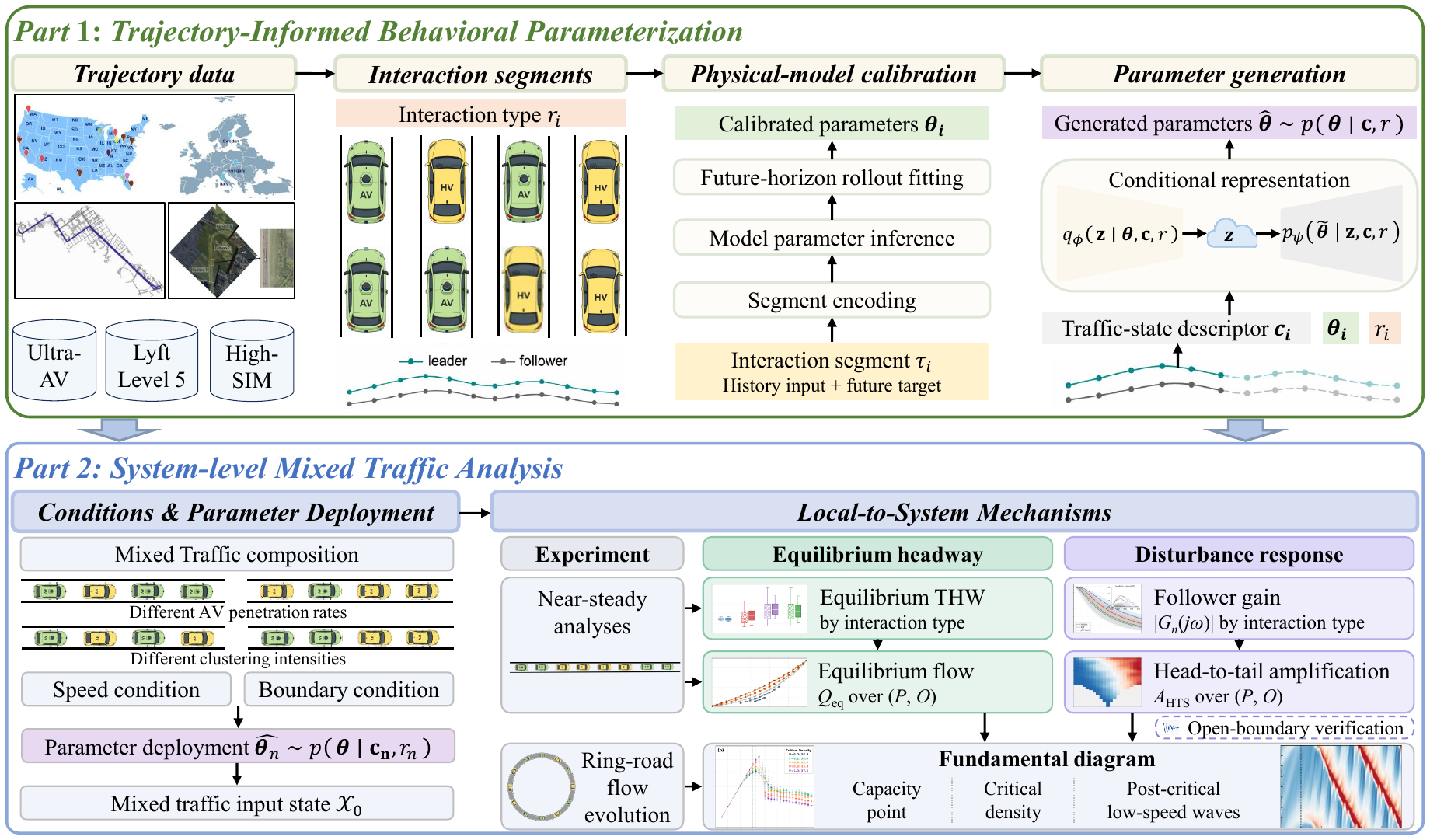}
    \caption{
    The proposed overall framework. In Part 1, calibrated and generated parameters are obtained from HV/AV trajectories (refer to Sections~\ref{s4} and~\ref{s6}). In Part 2, these parameters are deployed under prescribed mixed traffic conditions (left), and equilibrium headway and disturbance response are traced through near-steady analyses to the fundamental diagram from the ring-road simulations (right), with open-boundary simulations verifying the analytical stability results (refer to Sections~\ref{s5} and~\ref{s7}).
    }
    \label{fig:overall_framework}
\end{figure}

This study develops a framework that links large-scale HV/AV trajectory evidence to system-level mixed traffic analysis through behavioral parameterization. The framework consists of two parts, as illustrated in Fig.~\ref{fig:overall_framework}.

The first part of the framework converts trajectory data into behavioral inputs. Short car-following interaction segments are extracted from large-scale HV/AV trajectories and used to calibrate car-following parameters for different interaction types, thereby forming a discrete empirical parameter set. These calibrated samples are then paired with corresponding short-term traffic-state descriptors and transformed into traffic-state-conditioned conditional parameter representations. The resulting representations provide empirically grounded and comparable behavioral inputs for the subsequent mixed traffic analyses.

The second part of the framework evaluates system-level mixed traffic outcomes under different traffic conditions. AV penetration rate and clustering intensity are jointly varied to characterize mixed traffic composition, and the resulting systems are analyzed under representative speed and boundary settings. Near-steady analyses are used to examine equilibrium flow at prescribed speeds and analytical stability. Periodic-boundary flow-evolution analyses are used to investigate fundamental diagram characteristics, critical density, and space--time evolution of congestion. Together, these analyses provide an integrated evaluation of mixed traffic evolution under different traffic compositions and traffic conditions.

\section{Trajectory-informed and traffic-state-conditioned behavioral parameterization\label{s4}}

\subsection{Two-stage formulation and base car-following model}

To support a comparable representation of local car-following behavior in mixed traffic and an analytically tractable analysis of system-level evolution, the proposed framework represents heterogeneous local interactions using a unified physics-based car-following model. The task is then to construct a traffic-state-conditioned parameterization of this common model from large-scale HV/AV trajectory data.

Based on this formulation, the proposed framework consists of two stages. In the first stage, each car-following interaction segment \(\tau_i\) with interaction type \(r_i\) is mapped to a calibrated parameter sample
\begin{equation}
\boldsymbol{\theta}_i=\mathcal{G}(\tau_i,r_i),
\label{eq:calibration_mapping}
\end{equation}
thereby forming a discrete empirical parameter set
\begin{equation}
\mathcal{S}_{\theta}=\{(\tau_i,r_i,\boldsymbol{\theta}_i)\}_{i=1}^{N_{\mathrm{seg}}},
\label{eq:discrete_param_set}
\end{equation}
where \(N_{\mathrm{seg}}\) denotes the total number of interaction segments retained after data extraction and sliding-window segmentation, as described in Section~\ref{s4.2}.

In the second stage, a short-term traffic-state descriptor is extracted from each interaction segment
\begin{equation}
\mathbf{c}_i=\Phi(\tau_i),
\label{eq:state_descriptor_general}
\end{equation}
and paired with the corresponding interaction type and calibrated parameter sample to form an empirical conditional training set. Based on these empirical samples, the discrete parameter set is used to learn a conditional parameter representation
\begin{equation}
p(\boldsymbol{\theta}\mid \mathbf{c},r),
\label{eq:cond_param_dist}
\end{equation}
which maps a given local traffic state \(\mathbf{c}\) and interaction type \(r\) to a continuous distribution of plausible parameter values.

Under this formulation, behavioral inputs for subsequent mixed traffic analysis are generated through conditional sampling
\begin{equation}
\hat{\boldsymbol{\theta}}\sim p(\boldsymbol{\theta}\mid \mathbf{c},r),
\label{eq:cond_sampling}
\end{equation}
which allows behavioral parameters to be supplied under specified local traffic states and interaction types, rather than being restricted to the finite set of observed empirical realizations.

The IDM~\citep{treiber2000congested} is selected as the base physical model because of its compact parameter structure, physical interpretability, and widespread use in mixed traffic analysis. Its acceleration function, denoted by \(f_{\mathrm{IDM}}\), is
\begin{equation}
\hat{a}_n(t)=a_{\max}\left[1-\left(\frac{v_n(t)}{v_0}\right)^{\delta}
-\left(\frac{s_n^{*}(t)}{s_n(t)}\right)^2\right],
\label{eq:ch4_idm_acceleration}
\end{equation}
where the desired dynamic spacing is
\begin{equation}
s_n^{*}(t)=s_0+\max\left\{0,\;v_n(t)T+\frac{v_n(t)\Delta v_n(t)}{2\sqrt{a_{\max}b}}\right\},
\label{eq:ch4_idm_desired_spacing}
\end{equation}
and
\begin{equation}
\Delta v_n(t)=v_n(t)-v_{n-1}(t).
\label{eq:ch4_relative_speed}
\end{equation}
Here, \(\hat{a}_n(t)\) denotes the modeled longitudinal acceleration of the following vehicle, while \(v_n(t)\), \(\Delta v_n(t)\), and \(s_n(t)\) denote its longitudinal velocity, relative velocity, and gap, respectively. The corresponding parameter vector is defined as
\begin{equation}
\boldsymbol{\theta}=[s_0,T,b,a_{\max},v_0,\delta],
\label{eq:ch4_idm_parameter_vector}
\end{equation}
where \(s_0\) is the minimum gap, \(T\) is the desired time headway, \(b\) is the comfortable deceleration, \(a_{\max}\) is the maximum acceleration, \(v_0\) is the desired velocity, and \(\delta\) is the acceleration exponent.

\subsection{Interaction-segment construction and calibration dataset\label{s4.2}}

Longitudinal car-following events are extracted and preprocessed across all data sources using the criteria adopted in Ultra-AV~\citep{zhou2024unified}.

Each retained car-following event is segmented into fixed-length short interaction segments. In this study, a 4~s observation window is adopted. This choice follows recent mixed traffic behavior analysis based on Lyft Level 5~\citep{tang2025comparing}, which reported that most driving primitives lie within 1--5~s and that the mean primitive durations across different car-following modes are approximately 3.46--4.13~s. To improve data utilization and obtain a sufficiently rich empirical parameter set, the segmentation is implemented using a sliding-window strategy. Each resulting segment is denoted by
\begin{equation}
\tau_i=\{v_n(t),a_n(t),\Delta v_n(t),s_n(t)\}_{t=1}^{T_{\tau}},
\label{eq:ch4_interaction_segment}
\end{equation}
where \(\tau_i\) denotes the \(i\)th interaction segment, \(T_{\tau}\) denotes the number of time steps in the segment, and \(a_n(t)\) denotes the observed acceleration sequence of the following vehicle.

Each segment is then assigned an interaction-type label and organized as one calibration sample. Following the four interaction types considered in this study, the calibration dataset is written as
\begin{equation}
\mathcal{D}_{\mathrm{cal}}=\{(\tau_i,r_i)\}_{i=1}^{N_{\mathrm{seg}}},
\label{eq:ch4_calibration_dataset}
\end{equation}
where \(r_i\) denotes the interaction type of segment \(i\). This dataset serves as the direct input to the subsequent interaction-specific parameter calibration.

\subsection{Interaction-specific parameter calibration}

\subsubsection{Calibration module and expert routing}

Based on the constructed interaction segments, an interaction-specific calibration module is used to infer physically interpretable IDM parameter samples from observed local car-following trajectories. This step provides the discrete empirical basis of heterogeneous follower--leader behavior under the unified physical model. The resulting calibrated parameter samples serve as the empirical reference for the traffic-state-conditioned conditional parameter representation developed in Section~\ref{s4.4}.

Given an interaction segment \(\tau_i\) and its interaction type \(r_i\), the calibration task is to infer the corresponding calibrated parameter sample \(\boldsymbol{\theta}_i\). In the present implementation, each sample is represented by a short \(H\)-step local interaction history and encoded by a hybrid backbone combining Mamba~\citep{gu2023mamba} and Transformer~\citep{vaswani2017attention} blocks. The encoded representation is passed to an interaction-type head, whose predicted probabilities are used to activate a soft top-\(k\) expert routing mechanism. Each selected expert outputs a candidate IDM parameter sample, and the routed candidates are evaluated through the IDM-embedded rollout objective defined below.

This design allows interaction-type information to guide calibration without assigning all samples of the same type to a single deterministic parameter mode. The interaction-specific calibration follows the mapping defined in Eq.~\eqref{eq:calibration_mapping}, where \(\mathcal{G}(\cdot)\) denotes the calibration module applied to labeled interaction segments.

\subsubsection{IDM-embedded rollout and calibration objective}

The inferred parameters are evaluated by comparing the velocity and gap time series simulated by the embedded IDM with the corresponding observations. Let \(\boldsymbol{\theta}^{(k)}_i\) denote the candidate parameter sample produced by expert \(k\) for segment \(i\), and let \(w_{i,k}\) denote its normalized routing weight within the selected expert set~\(\mathcal{K}_i\). The modeled longitudinal acceleration at time \(t\) is written as
\begin{equation}
\hat{a}_n(t)=\Pi_{[\underline{a},\,\overline{a}]}
\left(
\sum_{k\in\mathcal{K}_i} w_{i,k}\,
f_{\mathrm{IDM}}\!\left(v_n(t),\Delta v_n(t),s_n(t);\boldsymbol{\theta}^{(k)}_i\right)
\right),
\label{eq:ch4_weighted_idm_acceleration}
\end{equation}
where \(\Pi_{[\underline{a},\,\overline{a}]}(\cdot)\) enforces the acceleration limits. Positive acceleration is naturally bounded by \(a_{\max}\) for each expert and by the weighted mean of these limits for the routed acceleration; the lower bound limits excessive braking. Starting from the terminal state of the observed history, the embedded IDM performs an \(F\)-step rollout, where \(F\) denotes the rollout horizon, to reproduce the future velocity and gap evolution, yielding \(\hat{v}_n(t)\) and \(\hat{s}_n(t)\). At each step, the follower velocity is constrained to be non-negative before updating the gap. The updated states are then used in the next step with the observed leader velocity as input. 

The calibration objective combines trajectory reconstruction, interaction-type discrimination, and physical regularization, according to
\begin{equation}
\mathcal{L}_{\mathrm{cal}}
=
\mathcal{L}_{\mathrm{traj}}
+
\mathcal{L}_{\mathrm{cls}}
+
\eta\,\mathcal{L}_{\mathrm{phy}},
\label{eq:ch4_calibration_loss}
\end{equation}
where
\begin{equation}
\mathcal{L}_{\mathrm{traj}}
=
\mathcal{L}_{v}+\mathcal{L}_{s}
=
\frac{1}{F}\sum_{t=1}^{F}\bigl(\hat{v}_n(t)-v_n(t)\bigr)^2
+
\frac{1}{F}\sum_{t=1}^{F}\bigl(\hat{s}_n(t)-s_n(t)\bigr)^2,
\label{eq:ch4_trajectory_loss}
\end{equation}
measures the discrepancy between the reproduced and observed future trajectories, and
\begin{equation}
\mathcal{L}_{\mathrm{cls}}
=
-\sum_{m=1}^{4} y_{i,m}\log \hat{p}_{i,m},
\label{eq:ch4_classification_loss}
\end{equation}
is the cross-entropy loss for interaction-type prediction, with \(\hat{p}_{i,m}\) denoting the predicted probability of type \(m\) and \(y_{i,m}\) the corresponding one-hot label. During training, the coefficient \(\eta\) is adaptively adjusted according to the relative scale between the main fitting loss and the physical penalty, so that physical regularization remains effective without overwhelming the trajectory-fitting objective.

\subsubsection{Physical regularization and calibrated parameter set}

During training, the routed acceleration in Eq.~\eqref{eq:ch4_weighted_idm_acceleration} is used to stabilize the trajectory reconstruction loss \(\mathcal{L}_{\mathrm{traj}}\). For each interaction segment, the candidate parameter sample associated with the highest routing weight is retained as the calibrated IDM parameter sample. Specifically, with
\begin{equation}
k_i^{*}=\arg\max_{k\in\mathcal{K}_i} w_{i,k},
\label{eq:ch4_selected_expert}
\end{equation}
the retained calibrated parameter sample is
\begin{equation}
\boldsymbol{\theta}_i=\boldsymbol{\theta}^{(k_i^{*})}_i.
\label{eq:ch4_representative_parameter}
\end{equation}

Physical regularization is imposed on the retained parameter sample to keep the calibrated IDM parameters within the feasible domain. Denote the lower and upper bounds by
\begin{equation}
\boldsymbol{\theta}_{\mathrm{L}}
=
[s_{0,\mathrm{L}},T_{\mathrm{L}},b_{\mathrm{L}},a_{\max,\mathrm{L}},v_{0,\mathrm{L}},\delta_{\mathrm{L}}],
\qquad
\boldsymbol{\theta}_{\mathrm{U}}
=
[s_{0,\mathrm{U}},T_{\mathrm{U}},b_{\mathrm{U}},a_{\max,\mathrm{U}},v_{0,\mathrm{U}},\delta_{\mathrm{U}}].
\label{eq:theta_bounds}
\end{equation}
The physical loss is written as
\begin{equation}
\mathcal{L}_{\mathrm{phy}}
=
\sum_{j=1}^{J}\xi_j
\left[
\mathrm{ReLU}(\theta_{\mathrm{L},j}-\theta_{i,j})
+
\mathrm{ReLU}(\theta_{i,j}-\theta_{\mathrm{U},j})
\right],
\label{eq:calib_phy_loss}
\end{equation}
where \(J=6\) denotes the number of IDM parameters, \(\theta_{i,j}\) denotes the \(j\)th component of \(\boldsymbol{\theta}_i\), and \(\xi_j\) denotes the parameter-specific regularization weight. The detailed parameter bounds are given in Table~\ref{tab:validation_settings}.

Applying the above procedure to all samples in \(\mathcal{D}_{\mathrm{cal}}\) yields the discrete empirical parameter set in Eq.~\eqref{eq:discrete_param_set}. This set contains the calibrated IDM parameter samples used in the next subsection to learn the traffic-state-conditioned conditional parameter distribution in Eq.~\eqref{eq:cond_param_dist}.

\subsection{Traffic-state characterization and conditional parameter representation\label{s4.4}}

\subsubsection{Traffic-state descriptor and conditional training samples}

The calibrated parameter samples in Eq.~\eqref{eq:discrete_param_set} provide a discrete empirical representation of interaction-specific car-following behavior. To deploy these behavioral inputs under prescribed local traffic states in subsequent mixed traffic analyses, the discrete empirical set is transformed into a traffic-state-conditioned conditional parameter representation.

For each calibrated interaction segment, a short-term traffic-state descriptor is constructed from local kinematic statistics. Following Eq.~\eqref{eq:state_descriptor_general}, the descriptor used in the present study is specified as
\begin{equation}
\mathbf{c}_i=
[\bar{v}_i,\bar{a}_i,\overline{\Delta v}_i,\overline{\mathrm{THW}}_i]^{\top}.
\label{eq:state_descriptor_inst}
\end{equation}
Here, \(\bar{v}_i\), \(\bar{a}_i\), \(\overline{\Delta v}_i\), and \(\overline{\mathrm{THW}}_i\) denote the mean velocity, mean acceleration, mean relative velocity, and mean time headway (THW), respectively, extracted from the interaction segment. The interaction type \(r_i\) is encoded as a one-hot vector and combined with \(\mathbf{c}_i\) as the conditional input to the generative module.

Based on the calibrated parameter samples, the corresponding traffic-state descriptors, and the interaction types, the empirical conditional training set is written as
\begin{equation}
\mathcal{D}_{\mathrm{cp}}=\{(\mathbf{c}_i,r_i,\boldsymbol{\theta}_i)\}_{i=1}^{N_{\mathrm{seg}}}.
\label{eq:condition_parameter_pairs}
\end{equation}
These empirical condition--parameter pairs are used to learn the conditional parameter distribution defined in Eq.~\eqref{eq:cond_param_dist}. 

\subsubsection{Conditional generative parameter representation}

To learn the conditional parameter distribution \(p(\boldsymbol{\theta}\mid\mathbf{c},r)\), a conditional generative module is adopted and implemented as a Conditional Variational Autoencoder (CVAE)~\citep{sohn2015learning}. During training, the encoder maps the calibrated parameter sample, the traffic-state descriptor, and the interaction type into a latent variable distribution \(q_{\phi}(\mathbf{z}\mid \boldsymbol{\theta},\mathbf{c},r)\). The decoder then reconstructs a parameter sample from the latent variable and the conditional input through \(p_{\psi}(\tilde{\boldsymbol{\theta}}\mid \mathbf{z},\mathbf{c},r)\), where~\(\tilde{\boldsymbol{\theta}}\) denotes the reconstructed parameter sample. Through this variational formulation, the model learns a continuous traffic-state-conditioned parameter representation rather than a finite lookup table of empirical samples.

\subsubsection{Training objective and parameter deployment}

The training objective combines parameter reconstruction, latent regularization, and physical regularization:
\begin{equation}
\mathcal{L}_{\mathrm{rep}}
=
\mathcal{L}_{\mathrm{rec}}
+
\beta \mathcal{L}_{\mathrm{KL}}
+
\gamma \mathcal{L}_{\mathrm{phy}}^{\mathrm{rep}},
\label{eq:representation_loss}
\end{equation}
where \(\mathcal{L}_{\mathrm{rec}}=\|\tilde{\boldsymbol{\theta}}-\boldsymbol{\theta}\|_2^2\) measures the reconstruction error and
\begin{equation}
\mathcal{L}_{\mathrm{KL}}=
D_{\mathrm{KL}}\!\left(
q_{\phi}(\mathbf{z}\mid \boldsymbol{\theta},\mathbf{c},r)
\;\|\;
\mathcal{N}(\mathbf{0},\mathbf{I})
\right)
\label{eq:ch4_kl_loss}
\end{equation}
is the Kullback--Leibler (KL) divergence between the latent posterior and the standard Gaussian prior. The physical regularization term \(\mathcal{L}_{\mathrm{phy}}^{\mathrm{rep}}\) is defined in Eq.~\eqref{eq:rep_phy_loss}.

The physical regularization adopts the same feasible parameter domain introduced in Eq.~\eqref{eq:theta_bounds}. Its bound-violation term follows the same form as Eq.~\eqref{eq:calib_phy_loss}, but is applied here to the reconstructed parameter sample:
\begin{equation}
\mathcal{L}_{\mathrm{bound}}^{\mathrm{rep}}
=
\sum_{j=1}^{J}\xi_j
\left[
\mathrm{ReLU}(\theta_{\mathrm{L},j}-\tilde{\theta}_j)
+
\mathrm{ReLU}(\tilde{\theta}_j-\theta_{\mathrm{U},j})
\right].
\label{eq:rep_bound_loss}
\end{equation}
Finally, the resulting physical regularization is defined as
\begin{equation}
\mathcal{L}_{\mathrm{phy}}^{\mathrm{rep}}
=
\mathcal{L}_{\mathrm{bound}}^{\mathrm{rep}}
+
\lambda_v \mathcal{L}_{v_0}^{\mathrm{rep}},
\label{eq:rep_phy_loss}
\end{equation}
with
\begin{equation}
\mathcal{L}_{v_0}^{\mathrm{rep}}
=
\mathrm{ReLU}(\bar{v}-\tilde{v}_0),
\label{eq:ch4_v0_loss}
\end{equation}
where \(\bar{v}\) denotes the local mean velocity contained in the traffic-state descriptor defined in Eq.~\eqref{eq:state_descriptor_inst}. This term is evaluated in physical units to prevent the reconstructed desired velocity from falling below the conditioned local velocity level during training.

Once trained, the conditional representation is used for parameter deployment in subsequent mixed traffic simulations. During deployment, the encoder is discarded. Given a specified traffic-state descriptor \(\mathbf{c}\) and interaction type \(r\), a latent variable is sampled from the standard Gaussian prior, \(\mathbf{z}\sim \mathcal{N}(\mathbf{0},\mathbf{I})\), and the decoder generates a behavioral parameter sample. For consistency with the deployment interface in Eq.~\eqref{eq:cond_sampling}, this generated parameter sample is denoted as
\begin{equation}
\hat{\boldsymbol{\theta}}
\sim
p_{\psi}(\boldsymbol{\theta}\mid \mathbf{z},\mathbf{c},r).
\label{eq:ch4_deployment_sampling}
\end{equation}

\section{Experimental setup\label{s5}}
\subsection{Mixed traffic state construction and parameter deployment\label{s5.1}}

In this study, the composition of mixed traffic is characterized by the AV penetration rate \(P\) and the clustering intensity \(O\), where the latter describes the organization of AVs and HVs within the vehicle sequence.

Let \(A_n\in\{0,1\}\) be a binary random variable representing the type of vehicle \(n\) in a single-lane vehicle sequence of length \(N\), where \(A_n=1\) represents an AV and \(A_n=0\) represents an HV. The AV penetration rate is defined as
\begin{equation}
P=\mathbb{E}\!\left(\frac{1}{N}\sum_{n=1}^{N}A_n\right),
\label{eq:ch5_penetration}
\end{equation}
where the expectation is taken over randomly generated vehicle sequences.

Given a penetration rate \(P\) and a feasible clustering intensity \(O\), the vehicle sequence is generated sequentially through a two-state Markov chain~\citep{ghiasi2017mixed, yao2023analysis, li2026unveiling}. The type of the first vehicle is sampled from
\begin{equation}
\Pr(A_1=1)=P,\qquad \Pr(A_1=0)=1-P,
\label{eq:ch5_initial_distribution}
\end{equation}
and the remaining vehicle types are generated according to
\begin{equation}
\Pr(A_{n+1}=j\mid A_n=i)=t_{ij}(P,O), \qquad i,j\in\{0,1\}.
\label{eq:ch5_transition_prob}
\end{equation}
The transition probabilities are specified as
\begin{equation}
\begin{aligned}
t_{11} &= P+O(1-P), &\qquad t_{10} &= (1-O)(1-P),\\
t_{01} &= (1-O)P,   &\qquad t_{00} &= (1-P)+OP.
\end{aligned}
\label{eq:ch5_transition_entries}
\end{equation}
These transition probabilities ensure that the pair \((P,O)\) determines not only the expected AV share but also the stochastic ordering of vehicle types in the initialized mixed traffic state.

The clustering intensity satisfies \(O\in[-1,1]\). When \(O=0\), the probability of the following vehicle type is independent of the type of its preceding vehicle, so the sequence approaches random mixing. When \(O\) increases, adjacent vehicles of the same type become more likely and the sequence becomes more clustered. When \(O\) decreases, alternating arrangements become more likely. However, this full interval is not feasible for every \(P\). For a given penetration rate, the transition probabilities must remain within \([0,1]\) while preserving \(P\) as the expected AV proportion. For \(0<P<1\), the admissible range of \(O\) depends on \(P\), and is given by
\begin{equation}
O\in [O_{\min}(P),\,1], \qquad
O_{\min}(P)=\max\!\left(-\frac{P}{1-P},\,-\frac{1-P}{P}\right).
\label{eq:ch5_O_feasible}
\end{equation}
This constraint also explains why highly alternating arrangements cannot always be attained when AVs and HVs are strongly unbalanced in proportion. For the pure HV and pure AV cases, the clustering intensity \(O\) is immaterial because only one vehicle type is present.

Once the vehicle type realization \(\{A_n\}_{n=1}^{N}\) is generated, the composition of follower--leader interaction types follows directly from adjacent vehicle pairs. Let vehicle \(n\) follow vehicle \(n-1\). The interaction type of follower \(n\) is then defined as
\begin{equation}
r_n=\Psi(A_n,A_{n-1}), \qquad n=2,\dots,N,
\label{eq:ch5_interaction_map}
\end{equation}
where \(\Psi(\cdot)\) maps each follower--leader pair into one of the four interaction types considered in this study. The empirical share of interaction type \(r\) in one initialized state is computed as
\begin{equation}
\rho_r=\frac{1}{N-1}\sum_{n=2}^{N}\mathbb{I}(r_n=r), \qquad r\in\mathcal{R},
\label{eq:ch5_interaction_share}
\end{equation}
where \(\mathcal{R}\) denotes the set of the four interaction types and \(\mathbb{I}(\cdot)\) is the indicator function. Therefore, once \(P\) and \(O\) are specified, the interaction-type composition follows directly from the generated mixed traffic state.

For the simulation settings considered in this study, the initial mixed traffic state is constructed under near-steady-state conditions. This treatment is adopted because the subsequent analyses start from mixed traffic states that are intended to represent relatively consistent car-following conditions across vehicles. Accordingly, for each follower, a near-steady local traffic state is constructed under the prescribed initial speed condition and then used for parameter deployment.

Under this initialization, the acceleration and relative velocity components are set to zero. When the initial time headway is not explicitly specified by the experiment setting, the time headway component is sampled from the empirical distribution of near-steady segments extracted from the trajectory dataset at the corresponding initial speed. Thus, for follower \(n\), the local traffic state is written as
\begin{equation}
\mathbf{c}_n=\left[v_n,\;0,\;0,\;\mathrm{THW}_n\right]^{\top},
\qquad n=2,\dots,N,
\label{eq:ch5_local_state}
\end{equation}
where \(v_n\) denotes the initial velocity of follower \(n\), and \(\mathrm{THW}_n\) denotes the initial time headway. Given the constructed local traffic state \(\mathbf{c}_n\) and the interaction type \(r_n\), the follower-specific behavioral parameters are sampled as
\begin{equation}
\hat{\boldsymbol{\theta}}_n \sim p(\boldsymbol{\theta}\mid \mathbf{c}_n,r_n),
\qquad n=2,\dots,N,
\label{eq:ch5_parameter_deployment}
\end{equation}
where \(\hat{\boldsymbol{\theta}}_n\) serves as the behavioral input for follower \(n\) in the subsequent mixed traffic analysis. This deployment follows directly from the conditional parameter representation defined in Section~\ref{s4.4}.

Accordingly, the mixed traffic input state for subsequent experiments can be summarized as
\begin{equation}
\mathcal{X}_0=
\left(
\{A_n\}_{n=1}^{N},\;
\{r_n,\mathbf{c}_n,\hat{\boldsymbol{\theta}}_n\}_{n=2}^{N}
\right),
\label{eq:ch5_state_summary}
\end{equation}
which combines the vehicle-type realization, the interaction-type composition, the initialized local traffic states, and the deployed behavioral parameters, thereby providing the common behavioral and compositional input for the subsequent system-level experiments under different traffic conditions.

\subsection{Experimental design\label{s5.2}}

\subsubsection{Traffic conditions and experiment families}

After the mixed traffic input state \(\mathcal{X}_0\) is constructed, the system-level analyses are organized by three traffic-condition dimensions: speed condition, mixed traffic composition, and roadway or boundary setting. Unless otherwise noted, the low-, medium-, and high-speed conditions correspond to initial speeds of 10, 20, and 30~m/s, respectively. The mixed traffic composition is specified by the AV penetration rate \(P\) and the clustering intensity \(O\), while the roadway or boundary setting determines whether the same behavioral inputs are evaluated under near-steady, open-boundary, or periodic-boundary settings.

Accordingly, three experiment families are considered: i)~near-steady analyses, to evaluate equilibrium flow and analytical stability around relatively stable operating states; ii)~open-boundary perturbation simulations, to verify the analytical stability results in the time domain under imposed sinusoidal leader disturbances; and~iii)~periodic-boundary ring-road flow-evolution analyses, to evaluate density--flow relationships, the critical density associated with the capacity point, and space--time speed evolution after the post-critical transition.

\subsubsection{Experiment-specific initialization and repeated realizations}

The initialization procedure depends on the experiment family. For near-steady analyses and open-boundary perturbation simulations, each follower starts from a near-steady local traffic state constructed under the prescribed speed condition. When the initial THW is not explicitly specified by the experiment setting, the THW component is sampled from the empirical near-steady THW distribution at the corresponding speed condition and used for parameter deployment. After the follower-specific parameters are deployed, the initial gap in simulation-based analyses is set to the equilibrium gap resulting from the deployed IDM parameters:
\begin{equation}
s_n^{\mathrm{eq}}=
\frac{\hat{s}_{0,n}+v\hat{T}_n}
{\sqrt{1-\left(v/\hat{v}_{0,n}\right)^{\hat{\delta}_n}}},
\label{eq:ch5_equilibrium_spacing}
\end{equation}
where \(\hat{s}_{0,n}\), \(\hat{T}_n\), \(\hat{v}_{0,n}\), and \(\hat{\delta}_n\) are the corresponding components of \(\hat{\boldsymbol{\theta}}_n\). This calculation is applied to feasible equilibrium samples for which the denominator in Eq.~(\ref{eq:ch5_equilibrium_spacing}) is well defined. The resulting gap allows the open-boundary simulations and near-steady analytical metrics to be initialized around the equilibrium state implied by the deployed behavioral parameters.

For periodic-boundary ring-road analyses, the initial spacing is determined by the prescribed target density rather than by Eq.~(\ref{eq:ch5_equilibrium_spacing}). The local traffic state used for parameter deployment keeps the prescribed speed condition, zero acceleration, and zero relative velocity components, while the THW component is determined by the density-conditioned initialization. 

Because both the vehicle ordering under \((P,O)\) and the follower-specific behavioral parameters are stochastic, each traffic condition is evaluated through repeated realizations. The resulting outputs are aggregated across realizations to obtain condition-level estimates of the corresponding equilibrium-flow, capacity, stability, and traffic-evolution metrics.

\subsubsection{Unified workflow of experiment design}

Algorithm~\ref{alg:ch5_workflow} summarizes the implementation workflow used across the three experiment families. 
The detailed evaluation metrics for the corresponding experiment families are elaborated in Section~\ref{s5.3}.

\begin{algorithm}
\caption{Unified workflow of experiment design under prescribed traffic conditions}
\label{alg:ch5_workflow}
\begin{algorithmic}[1]
\Require Traffic condition specified by the speed condition, the penetration rate \(P\), the clustering intensity \(O\), and the roadway or boundary setting
\Ensure Aggregated system-level metrics under the prescribed traffic condition

\For{each random realization under the prescribed traffic condition}
    \State Generate a vehicle type sequence \(\{A_n\}_{n=1}^{N}\) from the prescribed \((P,O)\)
    \State Determine the interaction type \(r_n\) for each follower \(n=2,\dots,N\)
    \State Construct the local traffic state \(\mathbf{c}_n\) for each follower
    \State Deploy follower-specific behavioral parameters by sampling \(\hat{\boldsymbol{\theta}}_n \sim p(\boldsymbol{\theta}\mid \mathbf{c}_n,r_n)\), \(n=2,\dots,N\)
    \State Set the initial gaps from the IDM equilibrium relation or the prescribed ring-road density
    \State Form the mixed traffic input state \(\mathcal{X}_0\)
    \State Run the designated analytical or simulation-based procedure
    \State Record the resulting system-level outputs
\EndFor

\State Aggregate the outputs across realizations
\State Compute the corresponding system-level metrics
\end{algorithmic}
\end{algorithm}

\subsection{System-level evaluation metrics\label{s5.3}}

The system-level metrics follow the three experiment families introduced in Section~\ref{s5.2}: equilibrium-flow and analytical stability metrics for the near-steady analyses, the simulation-based response gain for the open-boundary verification, and density--flow, critical-density, and space--time metrics for the periodic-boundary ring-road analyses. Detailed calculations and aggregation procedures are provided in Appendix~\ref{app:system_metrics}.

\subsubsection{Equilibrium-flow and stability metrics\label{s5.3.2}}

Equilibrium flow at the prescribed speed is evaluated using the equilibrium headways implied by the deployed IDM parameters~\citep{treiber2000congested}. The type-specific mean headway \(\bar{h}_r^{\mathrm{eq}}(v)\) is obtained by averaging \((s_n^{\mathrm{eq}}+l_{\mathrm{veh}})/v\) over followers of interaction type \(r\) at the prescribed speed \(v\), where \(s_n^{\mathrm{eq}}\) is obtained from Eq.~\eqref{eq:ch5_equilibrium_spacing} and \(l_{\mathrm{veh}}\) denotes the vehicle length. The detailed headway calculations are given in Appendix~\ref{app:metric_headways}. Following the headway-based capacity formulation for Markov-chain mixed traffic~\citep{ghiasi2017mixed}, the equilibrium flow under penetration rate\(P\), clustering intensity \(O\), and prescribed speed \(v\) is computed as
\begin{equation}
Q_{\mathrm{eq}}(P,O,v)
=
\frac{3600}
{\sum_{r\in\mathcal{R}}
\bar{\rho}_r(P,O)\bar{h}_{r}^{\mathrm{eq}}(v)} ,
\label{eq:ch5_capacity}
\end{equation}
where \(\bar{\rho}_r(P,O)\) denotes the expected interaction-type share induced by the transition probabilities. The headway is measured in seconds, so \(Q_{\mathrm{eq}}(P,O,v)\) is expressed in vehicles per hour per lane. For finite sampled mixed traffic states, \(\bar{\rho}_r(P,O)\) is estimated by the empirical share \(\rho_r\) defined in Eq.~(\ref{eq:ch5_interaction_share}), and the resulting equilibrium-flow estimates are aggregated over repeated realizations. This metric represents flow at the prescribed speed, rather than the maximum over speeds or densities. Ring-road capacity is identified separately from the peak of the simulated density--flow relationship in Section~\ref{s5.3.4}.

Analytical stability is first examined at the single-follower level. Following the standard linearization and frequency-domain treatment of car-following stability~\citep{feng_string_2019, montanino_string_2021}, \(G_n(j\omega)\) denotes the follower-level frequency response around the equilibrium state, where \(j\) is the imaginary unit. The linearization and transfer function are given in Appendix~\ref{app:metric_stability}. The maximum amplification of follower \(n\) and the corresponding dominant frequency are defined as
\begin{equation}
G_{n,\max}
=
\max_{\omega\in\Omega}
\left|G_n(j\omega)\right|,
\qquad
\omega_n^{*}
=
\arg\max_{\omega\in\Omega}
\left|G_n(j\omega)\right|,
\label{eq:ch5_single_vehicle_gain}
\end{equation}
where \(\Omega\) denotes the frequency grid used in the stability analysis. A value of \(G_{n,\max}>1\) indicates follower-level amplification at one or more frequencies. These metrics are summarized by interaction type and speed condition to characterize local stability tendencies implied by the deployed behavioral parameters.

The analysis then extends from individual followers to heterogeneous platoons. Since platoon-level disturbance transmission depends on the product of follower-level frequency responses~\citep{feng_string_2019, montanino_string_2021}, the head-to-tail amplification spectrum of the initialized mixed traffic input state \(\mathcal{X}_0\) is defined as
\begin{equation}
A_{\mathrm{HTS}}(\omega;\mathcal{X}_0)
=
\left|
\prod_{n=2}^{N}
G_n(j\omega)
\right|.
\label{eq:ch5_hts_spectrum}
\end{equation}
The maximum head-to-tail amplification and its dominant frequency are
\begin{equation}
A_{\mathrm{HTS}}^{\max}(\mathcal{X}_0)
=
\max_{\omega\in\Omega}
A_{\mathrm{HTS}}(\omega;\mathcal{X}_0),
\qquad
\omega_{\mathrm{HTS}}^{*}(\mathcal{X}_0)
=
\arg\max_{\omega\in\Omega}
A_{\mathrm{HTS}}(\omega;\mathcal{X}_0).
\label{eq:ch5_hts_max}
\end{equation}
To avoid classifying numerically neutral responses as unstable, a small tolerance \(\epsilon_A=0.001\) is used. The unstable frequency set \(\Omega_{\mathrm{unstable}}\) contains frequencies at which \(A_{\mathrm{HTS}}>1+\epsilon_A\), and a realization is classified as unstable when \(A_{\mathrm{HTS}}^{\max}>1+\epsilon_A\). For a traffic condition \(\kappa=(P,O,v)\), \(R_{\mathrm{unstable}}\) is the fraction of unstable cases over \(B\) repeated realizations. The corresponding indicator and aggregation formulas are provided in Appendix~\ref{app:metric_stability}.

Open-boundary simulations with sinusoidal leader perturbations are further used to verify the analytical stability results in the time domain; the imposed input, response amplitude, and simulation-based gain are defined in Appendix~\ref{app:metric_sinusoidal}.

\subsubsection{Periodic-boundary flow-evolution metrics\label{s5.3.4}}

Periodic-boundary flow-evolution metrics are used to evaluate density-conditioned mixed traffic dynamics in the ring-road simulations. These analyses focus on three related outputs: the aggregate density--flow relationship, the critical density associated with the capacity point, and the space--time organization of post-critical low-speed waves~\citep{zhou2020modeling, yao2022fundamental, sun2024fundamental}. The first two outputs describe the aggregate traffic-evolution pattern, whereas the space--time analysis illustrates how the post-critical transition develops dynamically within the periodic domain.

For each traffic condition \(\kappa=(P,O,v)\) and prescribed density \(\varrho\), vehicles are initialized on a single-lane ring road under periodic-boundary conditions and then evolved using the deployed behavioral parameters. A predefined warm-up period is first applied. Aggregate quantities are then computed over the statistical evaluation window \(\mathcal{T}_{\mathrm{stat}}\). For realization \(b\), \(\bar{v}^{(b)}(\varrho;\kappa)\) is the mean speed over all vehicles and sampled time steps in \(\mathcal{T}_{\mathrm{stat}}\), computed as detailed in Appendix~\ref{app:metric_ring}. The corresponding flow is calculated as
\begin{equation}
Q^{(b)}(\varrho;\kappa)
=
3.6\,\varrho\,\bar{v}^{(b)}(\varrho;\kappa),
\label{eq:ch5_fd_flow}
\end{equation}
where \(\varrho\) is measured in veh/km, \(\bar{v}^{(b)}\) is measured in m/s, and the factor 3.6 accounts for unit conversion. The condition-level mean speed and flow are obtained by aggregating \(\bar{v}^{(b)}(\varrho;\kappa)\) and \(Q^{(b)}(\varrho;\kappa)\) over repeated realizations.

The fundamental diagram is constructed from the aggregated density--flow relationship over the prescribed density grid \(\mathcal{D}_{\varrho}\). The mean flow \(\bar{Q}(\varrho;\kappa)\) is obtained by averaging \(Q^{(b)}(\varrho;\kappa)\) over \(B\) repeated realizations, as specified in Appendix~\ref{app:metric_ring}. The flow coordinate of the capacity point and the corresponding critical density are defined as
\begin{equation}
Q_{c}(\kappa)
=
\max_{\varrho\in\mathcal{D}_{\varrho}}
\bar{Q}(\varrho;\kappa),
\qquad
\varrho_{c}(\kappa)
=
\arg\max_{\varrho\in\mathcal{D}_{\varrho}}
\bar{Q}(\varrho;\kappa).
\label{eq:ch5_critical_density}
\end{equation}
Here, \(Q_c\) represents the peak flow in the aggregated density--flow relationship, while \(\varrho_c\) identifies the density location of the capacity point. The critical density is used to compare how the capacity point shifts across AV penetration rates, clustering intensities, and speed conditions. It also provides the density reference for selecting representative post-critical cases in the space--time analysis.

Space--time speed evolution is further examined under selected density conditions around or above \(\varrho_c\). For these cases, vehicle positions and speeds are recorded over time to characterize the emergence, persistence, and propagation of localized low-speed regions within the periodic domain. When organized low-speed bands are identifiable after the post-critical transition, their propagation speed is estimated from branch slopes in the space--time plane. The branch-based wave-speed estimate \(c_{\mathrm{wave}}\) is obtained by fitting the branch position against time, as detailed in Appendix~\ref{app:metric_ring}. Since the longitudinal coordinate \(x\) increases in the vehicle moving direction, a negative \(c_{\mathrm{wave}}\) indicates upstream propagation. These wave-speed estimates are used as auxiliary diagnostics of organized low-speed bands, while the main periodic-boundary flow-evolution metrics remain the fundamental diagram, the critical density, and the space--time speed patterns.

\section{Validation of behavioral parameterization\label{s6}}

\subsection{Validation settings and evaluation protocol\label{s6.1}}

This section validates the behavioral parameterization developed in Section~\ref{s4}. The validation examines three aspects: whether the calibrated IDM parameter samples reproduce observed short-horizon car-following evolution, whether the resulting discrete empirical parameter set captures interaction-specific heterogeneity, and whether the traffic-state-conditioned conditional parameter representation can provide generated behavioral inputs for microscopic mixed traffic simulation.

The calibration dataset is constructed from Ultra-AV~\citep{zhou2024unified}, HIGH-SIM~\citep{shi2021video}, and Lyft Level 5~\citep{mandal2020motion}. Ultra-AV is used as the primary standardized source of longitudinal AV car-following trajectories, providing data from open AV sources such as the Waymo Open Dataset~\citep{sun2020scalability}, Argoverse~2~\citep{wilson2023argoverse}, and OpenACC~\citep{makridis2021openacc}, under a unified longitudinal format and a consistent processing procedure. HIGH-SIM and Lyft Level 5 are further processed using the longitudinal extraction and preprocessing criteria adopted in Ultra-AV, so as to improve the coverage of the four interaction types and broaden the range of local traffic states represented in the retained interactions.

The evaluation uses the interaction-segment dataset constructed in Section~\ref{s4.2}. After sliding-window segmentation and filtering, the retained segment set contains approximately 2.01 million AV--AV, 2.42 million AV--HV, 2.70 million HV--AV, and 3.52 million HV--HV interaction segments. Calibration-stage reconstruction is evaluated on a subset that is not used to train the calibration module. The conditional representation is trained from the calibrated empirical condition--parameter pairs and monitored using an internal validation split.

Each 4~s interaction segment is divided into a 2~s historical window and a 2~s future rollout horizon, corresponding to \(H=20\) and \(F=20\) time steps at 10~Hz. The historical window is used for calibration-stage parameter inference or for constructing the short-term traffic-state descriptor used by the conditional representation, while the future horizon is used for rollout-based validation. The main numerical settings not specified in Section~\ref{s4} are summarized in Table~\ref{tab:validation_settings}.

For calibration-stage validation, the proposed calibration module is compared with conventional IDM calibration using a genetic algorithm (IDM-GA)~\citep{kesting2008calibrating} and three physics-informed parameter-inference baselines, PIDL-IDM, PILSTM-IDM~\citep{mo2021physics}, and PIT-IDM~\citep{geng2023physics}, each with sequential and joint training variants. The comparison uses velocity and gap root mean square error (RMSE) over the reported rollout horizons. For the conditional representation, reconstructed parameters on the held-out test set are compared with the calibrated samples, which serve as the empirical reference for \(p(\boldsymbol{\theta}\mid\mathbf{c},r)\), using moment statistics, kernel density comparisons, Kullback--Leibler divergence~\citep{kullback1951information}, and Jensen--Shannon (JS) divergence~\citep{lin2002divergence}. Generated parameters are then used in closed-loop IDM rollouts to assess short-horizon trajectory reconstruction, the distributions of velocity, gap, relative velocity, and time headway, and one-shot long-horizon extrapolation.

\subsection{Calibration accuracy and parameter heterogeneity\label{s6.2}}

Table~\ref{tab:calibration_rmse} reports the rollout-based reconstruction accuracy obtained from the calibrated IDM parameter samples. Overall, the proposed calibration module achieves the lowest average RMSE for both velocity and gap reconstruction. For velocity, IDM-GA performs best at the 0.5~s horizon, but its error increases more rapidly as the rollout horizon extends. The proposed module gives the lowest velocity RMSE from 1.0~s onward and the lowest average velocity~RMSE. For gap reconstruction, the proposed module gives the lowest RMSE across all reported horizons, indicating a stronger improvement in reproducing the longitudinal state evolution affected by accumulated velocity and acceleration differences.

The reconstructed gap and velocity values at the 2~s rollout endpoint are also concentrated near the observed values for all four interaction types (Fig.~\ref{fig:calibration_scatter}).


\begin{table}[pos=t]
\centering
\caption{Calibration-stage trajectory reconstruction accuracy and closed-loop rollout accuracy with generated parameters.}
\label{tab:calibration_rmse}
\fontsize{8.5}{10.5}\selectfont
\renewcommand{\arraystretch}{1.18}
\setlength{\tabcolsep}{4.2pt}
\begin{threeparttable}
\begin{tabular*}{\linewidth}{@{\extracolsep{\fill}}ll|ccccc|ccccc}
\toprule
\multicolumn{2}{l|}{\multirow{2}{*}{\textbf{Method}}}
& \multicolumn{5}{c|}{\textbf{Velocity RMSE (m/s)}}
& \multicolumn{5}{c}{\textbf{Gap RMSE (m)}} \\
\cmidrule(lr){3-7} \cmidrule(lr){8-12}
Category & Model
& 0.5s & 1.0s & 1.5s & 2.0s & Avg.
& 0.5s & 1.0s & 1.5s & 2.0s & Avg. \\
\midrule
Conventional
& IDM-GA
& \textbf{0.192} & 0.327 & 0.423 & 0.495 & 0.359
& 1.264 & 1.300 & 1.368 & 1.474 & 1.352 \\
\midrule
\multirow{6}{*}{Physics-informed}
& PIDL-IDM (Seq)
& 1.056 & 1.518 & 1.835 & 2.083 & 1.623
& 0.448 & 1.128 & 1.969 & 2.932 & 1.619 \\
& PIDL-IDM (Joint)
& 0.471 & 0.926 & 1.357 & 1.749 & 1.126
& 0.311 & 0.626 & 1.312 & 2.091 & 1.085 \\
& PILSTM-IDM (Seq)
& 0.645 & 1.272 & 1.838 & 2.311 & 1.517
& 0.344 & 0.886 & 1.697 & 2.752 & 1.420 \\
& PILSTM-IDM (Joint)
& 0.273 & 0.511 & 0.725 & 0.916 & 0.606
& 0.286 & 0.575 & 0.917 & 1.326 & 0.776 \\
& PIT-IDM (Seq)
& 0.568 & 0.691 & 1.071 & 1.818 & 1.037
& 0.292 & 0.491 & 0.877 & 1.525 & 0.796 \\
& PIT-IDM (Joint)
& 0.450 & 0.577 & 0.799 & 1.592 & 0.855
& 0.260 & 0.436 & 0.743 & 1.270 & 0.677 \\
\midrule
\multirow{2}{*}{Proposed}
& Calibration module
& 0.266 & \textbf{0.300} & \textbf{0.348} & \textbf{0.419} & \textbf{0.333}
& \textbf{0.214} & \textbf{0.343} & \textbf{0.466} & \textbf{0.613} & \textbf{0.409} \\
& Generated parameters
& 0.311 & 0.497 & 0.640 & 0.753 & 0.550
& 0.287 & 0.574 & 0.895 & 1.250 & 0.751 \\
\bottomrule
\end{tabular*}
\begin{minipage}{\linewidth}
\footnotesize
\textit{Note}: Seq and Joint denote sequential and joint training settings, respectively. The generated row reports closed-loop rollouts using IDM parameters produced by the learned conditional representation (Section~\ref{s6.3}); its calibrated reference coincides with the calibration-module row.
\end{minipage}
\end{threeparttable}
\end{table}

\begin{figure}[pos=t]
\centering
\includegraphics[width=\linewidth]{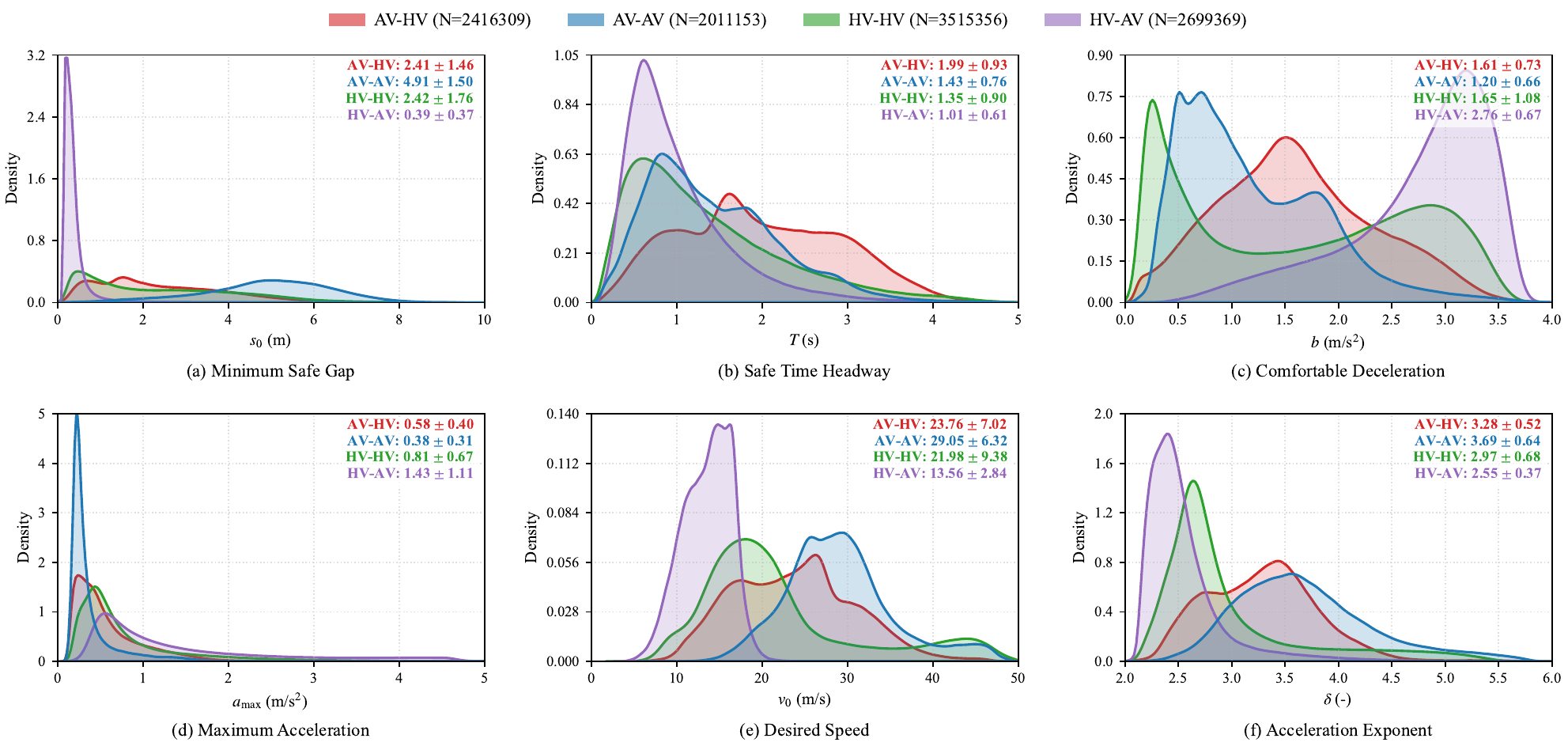}
\caption{Marginal distributions of calibrated IDM parameters by interaction type. The values shown in each panel indicate the mean and standard deviation of the corresponding calibrated parameter samples.}
\label{fig:calibrated_param_kde}
\end{figure}

After the rollout-based reconstruction accuracy is established, the calibrated IDM parameter samples are further examined as a discrete empirical parameter set. Figure~\ref{fig:calibrated_param_kde} shows the marginal distributions of the six calibrated IDM parameters for the four follower--leader interaction types. The distributions differ substantially across interaction types, indicating that a single common parameter distribution would be insufficient for representing heterogeneous car-following behavior in mixed traffic.
Most parameters show a clear follower-type pattern, and, except for the minimum gap \(s_0\), the distributions generally separate into AV-follower and HV-follower groups. The AV-follower interactions, including AV--AV and AV--HV, are associated with lower \(a_{\max}\), lower \(b\), higher \(v_0\), higher \(\delta\), and generally larger desired time headway \(T\) when pooled over all traffic states. By contrast, the HV-follower interactions, including HV--HV and HV--AV, show higher acceleration- and deceleration-related parameters, lower desired velocity values, and lower acceleration exponent values. This pattern suggests that the type of the following vehicle is a major source of calibrated parameter heterogeneity, while the leader type and local traffic state still affect the detailed distributional shape.
The main exception is \(s_0\), which shows stronger interaction-specific variation rather than a simple follower-type grouping. In particular, AV--AV interactions are concentrated at larger \(s_0\) values, whereas HV--AV interactions are concentrated near smaller values. This suggests that \(s_0\) is more closely related to the realized gap regime in the retained interaction segments and to the combined follower--leader pairing. The comfortable deceleration \(b\) follows the follower-type grouping in its central tendency, but the HV--HV distribution exhibits a visible bimodal shape. Further results shown in Appendix~\ref{app:b_distribution_by_dataset} indicate that this bimodality is closely related to the different operating environments represented by HIGH-SIM~\citep{shi2021video} and Lyft Level 5~\citep{mandal2020motion}. Therefore, the calibrated IDM parameter samples should be interpreted as trajectory-reproducing parameterizations that reflect both driving behavior and operating environment, rather than as isolated measurements of individual driving preferences.

\subsection{Generated behavior versus observation\label{s6.3}}

In reconstruction mode on the held-out test set, the conditional representation preserves the calibrated parameter distributions of all four interaction types, with joint Jensen--Shannon divergences of the six-dimensional parameter distributions below 0.017 (Table~\ref{tab:reconstructed_param_distribution}).

The generated IDM parameters are further evaluated through closed-loop rollouts over the same 2~s future horizon used in the calibration-stage validation. The overall results are reported in the generated-parameter row of Table~\ref{tab:calibration_rmse}, and the comparison with the calibrated reference for each interaction type is given in Table~\ref{tab:generated_rollout_rmse}. 
Replacing directly calibrated parameters with generated parameters generally increases the rollout error. This is expected because generated parameters are sampled from the learned conditional representation rather than fitted to each individual interaction segment. The increase is more evident for gap RMSE than for velocity RMSE, since gap evolution accumulates velocity and acceleration deviations during closed-loop rollout. Nevertheless, the generated-parameter rollouts remain stable across all four interaction types, with overall average RMSE values of 0.550~m/s for velocity and 0.751~m for gap. AV--AV shows the smallest errors, whereas interactions involving HVs show larger errors, especially in gap reconstruction.

\begin{figure}[pos=t]
\centering
\includegraphics[width=\linewidth]{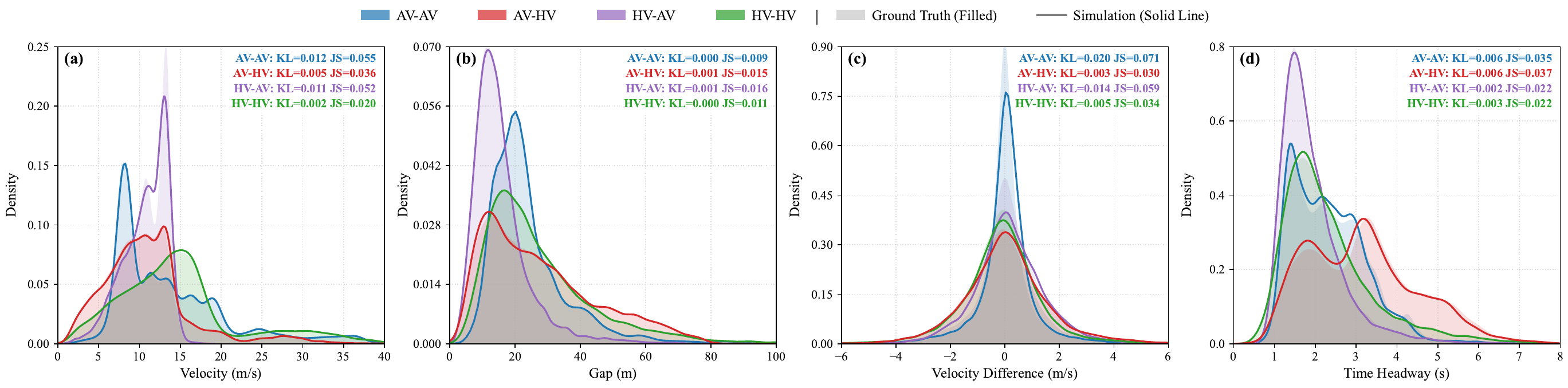}
\caption{Distributional consistency between observed and simulated trajectory-level indicators using generated IDM parameters. Filled densities denote observed values, and solid lines denote simulated values. KL (observed\,\(\|\)\,simulated) and JS values are computed from 49-bin histograms over the plotted ranges using natural logarithms; JS denotes the Jensen--Shannon distance.}
\label{fig:generated_indicator_distribution}
\end{figure}

Figure~\ref{fig:generated_indicator_distribution} further compares the observed and simulated distributions of trajectory-level indicators using generated parameters, including velocity, gap, relative velocity, and time headway. For visualization, 10,000 test segments are sampled from each interaction type.
The simulated distributions generally follow the observed distributions for the basic car-following variables and preserve the main type-specific patterns, whereas larger discrepancies appear mainly in the tails of the time-headway distributions. 

\begin{figure}[pos=t]
\centering
\includegraphics[width=\linewidth]{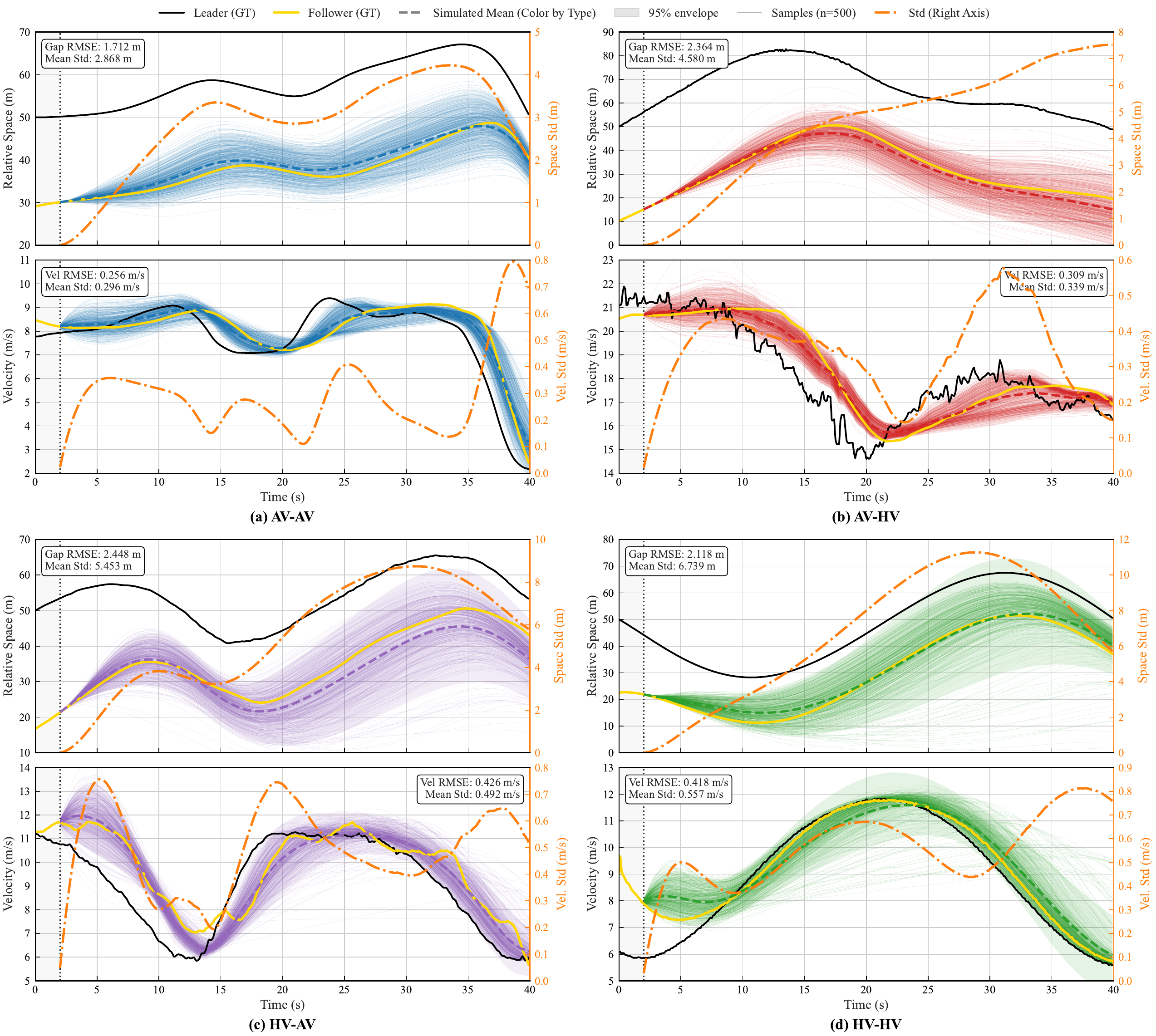}
\caption{One-shot 40~s closed-loop rollout examples using generated IDM parameters. The first 2~s window is used for condition construction, and the remaining horizon is simulated without re-calibration. Solid yellow lines denote observed follower trajectories, black lines denote observed leader trajectories, dashed colored lines denote the simulated mean, and shaded regions denote the 95\% simulation envelope.}
\label{fig:oneshot_long_horizon}
\end{figure}

The preceding validation focuses on short-horizon rollout accuracy and distributional consistency. To further examine whether generated parameters can support sustained closed-loop simulation, a one-shot 40~s rollout experiment is presented. For each interaction type, one illustrative test trajectory with relatively large velocity variation is selected. These cases are more demanding than near-steady following, because small parameter deviations are more likely to accumulate when the follower undergoes evident acceleration and deceleration changes. 
For each selected trajectory, the first 2~s historical window is used to construct the short-term traffic-state descriptor, and 500 IDM parameter samples are drawn from the learned conditional representation. These generated parameters are sampled once, kept fixed during the rollout, and used to simulate the follower for the remaining 38~s without re-calibration. The observed leader trajectory is used as the external leading-vehicle input, while the follower trajectory is simulated in closed loop.
For visual comparison, positions are plotted in a reference frame moving with the mean leader speed, whereas the reported gap RMSE is computed from the original gap series.
Figure~\ref{fig:oneshot_long_horizon} shows that the simulated mean trajectories follow the main evolution of the observed follower trajectories over the 40~s horizon. The velocity trajectories are reproduced with relatively small errors, with RMSE values ranging from about 0.26 to 0.43~m/s. The gap RMSE values are larger, ranging from about 1.71 to 2.45~m, which is expected because spatial deviations accumulate over the longer closed-loop rollout. Considering that the selected examples contain evident velocity variations and are simulated without re-calibration after the initial 2~s window, the results indicate that generated parameters can maintain stable long-horizon simulation beyond the short validation horizon.
The 95\% simulation envelopes further show the stochastic variation represented by the conditional parameter distribution. The AV--AV example has the narrowest envelope, whereas interactions involving HVs show broader dispersion, especially in position, which is consistent with the greater behavioral variability observed in HV-related interactions.

\section{Mixed traffic evolution: system-level results\label{s7}}

This section presents the system-level results obtained from the mixed traffic state construction and parameter deployment framework introduced in Section~\ref{s5}. The results are organized around two mechanisms through which interaction-specific car-following behavior shapes mixed traffic: the equilibrium headway, which governs flow efficiency, and the disturbance response, which governs stability. Sections~\ref{s7.1} and~\ref{s7.2} examine each mechanism from the interaction-type level to the system level, where AV penetration rate \(P\) and clustering intensity \(O\) determine how often and in what order the interaction types occur. Section~\ref{s7.3} then examines how the two mechanisms jointly shape the fundamental diagram in periodic-boundary ring-road simulations. Since both vehicle ordering and behavioral parameter deployment are stochastic, condition-level curves and maps are aggregated over repeated realizations unless otherwise noted. 

\subsection{From interaction-type headways to equilibrium flow\label{s7.1}}

\begin{figure}[pos=t]
\centering
\includegraphics[width=\linewidth]{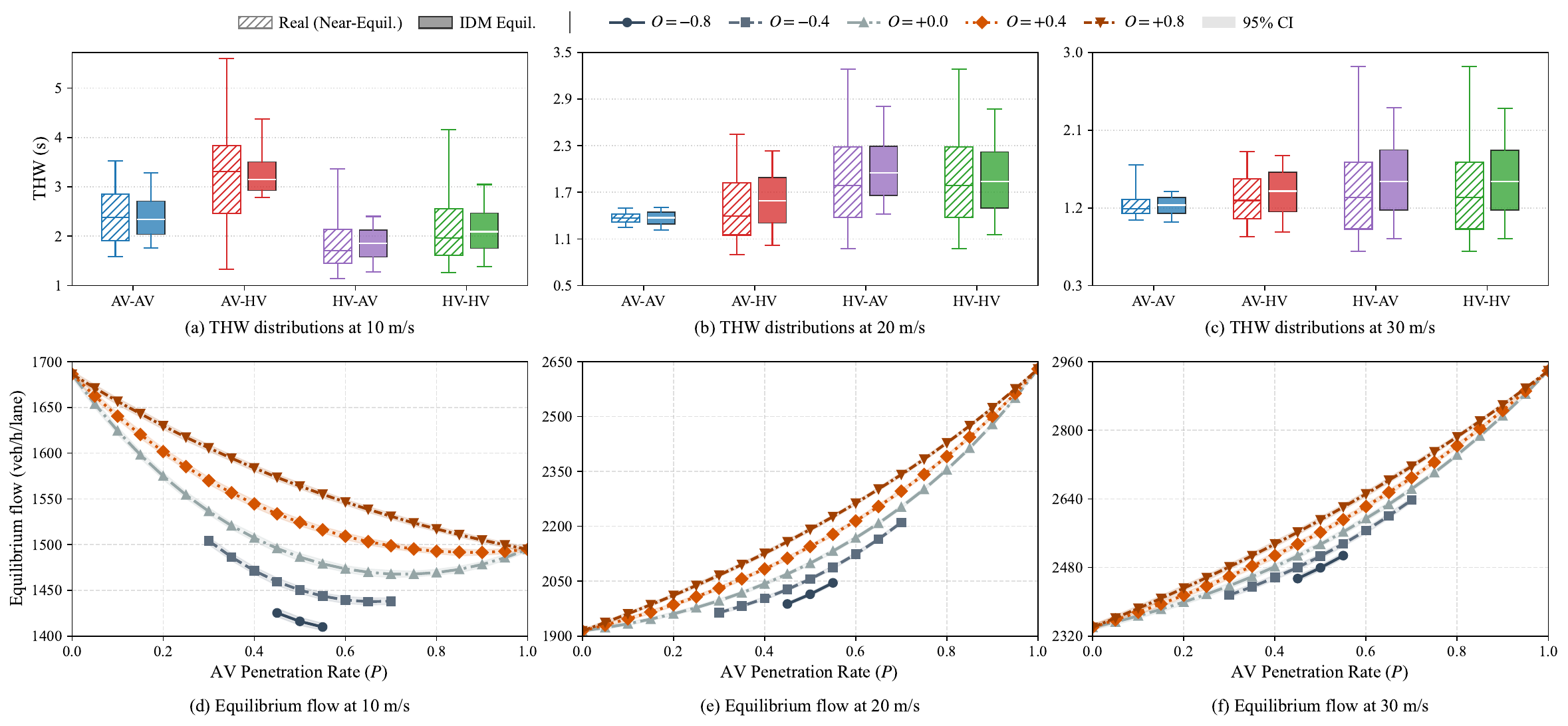}
\caption{Equilibrium time headways and equilibrium flow at prescribed speeds of 10, 20, and 30~m/s. Panels (a)--(c) compare empirical near-steady THW distributions (hatched) with IDM equilibrium THW distributions (solid) by interaction type. Panels (d)--(f) show mean equilibrium-flow estimates across AV penetration rate and clustering intensity; shaded bands denote 95\% bootstrap intervals across repeated parameter-deployment realizations, and only feasible \((P,O)\) combinations are plotted.}
\label{fig:capacity_variation}
\end{figure}

Figure~\ref{fig:capacity_variation} summarizes the equilibrium time headways and equilibrium-flow estimates under the three speed conditions. As shown in Fig.~\ref{fig:capacity_variation}(a)--(c), the IDM equilibrium THWs used for equilibrium-flow calculation remain broadly aligned with the empirical near-steady THW ranges, while retaining clear interaction-type differences. These differences reverse with speed. Under the 10~m/s condition, AV-follower interactions (AV--AV and AV--HV) have longer THWs than HV-follower interactions (HV--AV and HV--HV), whereas under the 20 and 30~m/s conditions, they have shorter THWs. Since equilibrium flow at a prescribed speed is determined by the weighted mean equilibrium headway, the equilibrium-flow patterns in Fig.~\ref{fig:capacity_variation}(d)--(f) mainly reflect how \(P\) and \(O\) change the frequency and ordering of follower--leader interaction types.

Moreover, under the 10~m/s condition, the pure-AV case yields a lower equilibrium flow than the pure-HV case because AV--AV interactions have larger equilibrium THWs than HV--HV interactions. At intermediate penetration rates, AV--HV interactions with relatively large equilibrium THWs increase the weighted mean equilibrium headway and further reduce equilibrium flow. Positive clustering partly mitigates this reduction by decreasing the share of AV--HV interactions.
The 20 and 30~m/s conditions show the opposite pattern. At these speeds, AV-follower interactions have shorter equilibrium THWs than HV-follower interactions, so higher \(P\) reduces the effective mean headway and increases equilibrium flow. For a fixed \(P\), larger positive \(O\) generally raises equilibrium flow at intermediate penetration rates by producing more same-type, especially AV-related, interaction sequences; this effect naturally weakens near the pure-HV and pure-AV endpoints.

\subsection{From follower-level gains to head-to-tail amplification\label{s7.2}}

\subsubsection{Follower-level stability characteristics by interaction type\label{s7.2.1}}

\begin{figure}[pos=t]
\centering
\includegraphics[width=\linewidth]{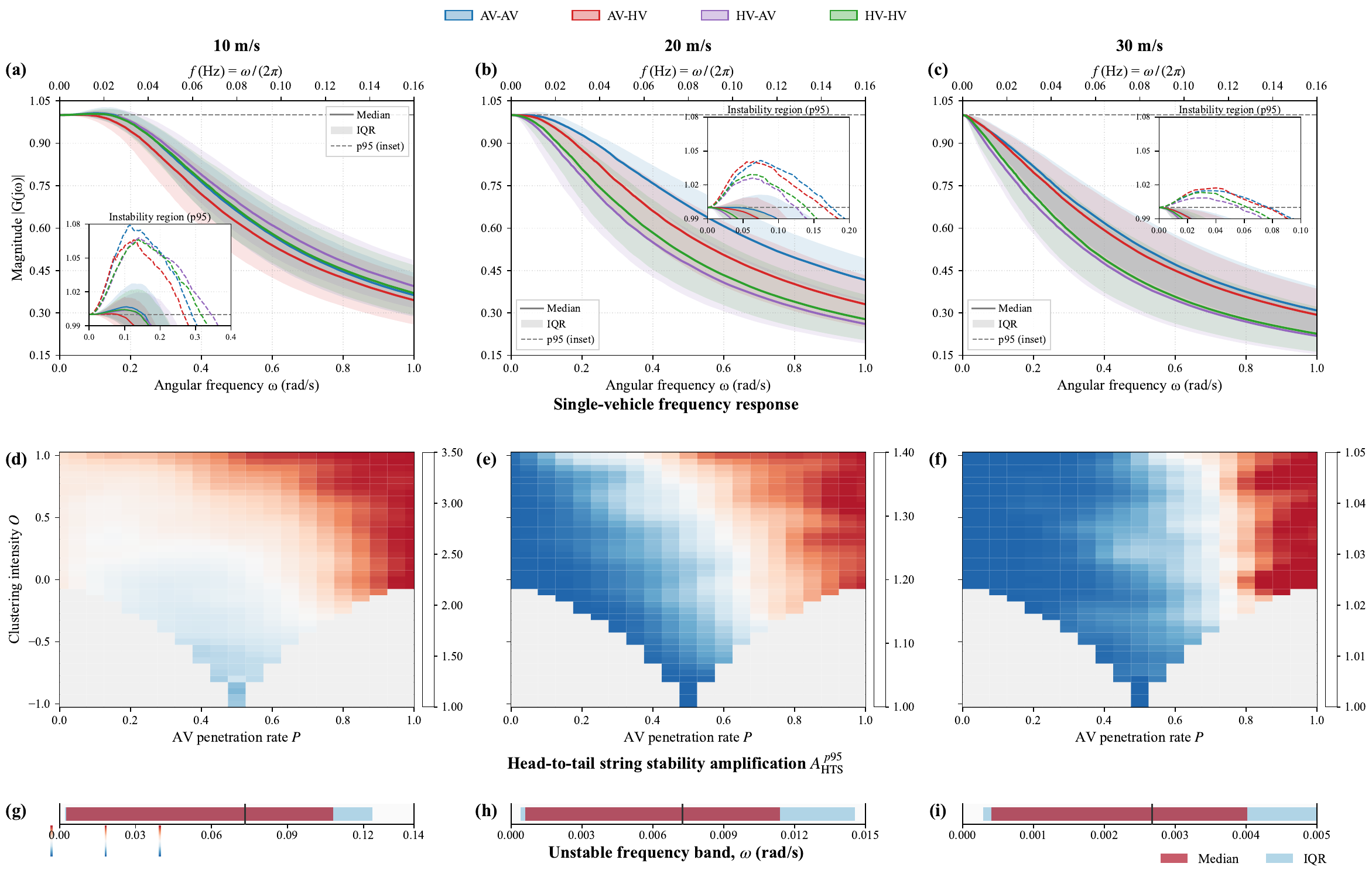}
\caption{Follower-level frequency responses and platoon-level head-to-tail amplification at equilibrium speeds of 10, 20, and $30~\mathrm{m/s}$ (left to right). Panels (a)--(c) show median follower-level responses and their interquartile ranges by interaction type; dashed curves in the insets show the 95th-percentile responses near the neutral amplification threshold. Panels (d)--(f) show the 95th percentile of the maximum head-to-tail amplification $A_{\mathrm{HTS}}^{\max}$; gray regions indicate infeasible $(P,O)$ combinations, and color scales are set separately for each speed. Panels (g)--(i) show the unstable frequency bands of unstable realizations pooled over all feasible compositions: red bars give the median band, blue bars the interquartile range of the band limits, and black lines the median dominant frequency $\omega_{\mathrm{HTS}}^*$. Platoon-level results use $B=100$ repeated realizations with $N=100$ vehicles per feasible composition, with instability defined by $A_{\mathrm{HTS}}^{\max}>1.001$.}
\label{fig:follower_level_stability}
\end{figure}

Figure~\ref{fig:follower_level_stability}(a)--(c) reports the follower-level frequency-response characteristics for the four follower--leader interaction types under the three equilibrium speed conditions. Overall, the responses remain close to the neutral amplification threshold, indicating that the local amplification implied by the deployed parameters is generally mild in magnitude.

The main distinction is speed dependent. Under the 10~m/s condition, several interaction types show slight low-frequency amplification, and the share of samples with \(G_{\max}>1\) is comparatively high (Fig.~\ref{fig:stability_supplement}(b)). Instead, under the 20 and 30~m/s conditions, the median responses are mostly non-amplifying, and amplification appears mainly in the upper tail of the sampled parameter distribution. The amplified frequency range also becomes narrower and shifts toward lower frequencies as the equilibrium speed increases.

Across interaction types, AV-follower interactions tend to retain larger upper-tail amplification than HV-follower interactions in the 20 and 30~m/s conditions. These follower-level results should therefore be interpreted as local response characteristics rather than direct mixed-flow stability outcomes. Their system-level effect depends on how often the corresponding interaction types occur and whether they occur consecutively along the vehicle sequence, which is examined through the platoon-level head-to-tail amplification analysis in the next subsection.

\subsubsection{Platoon-level head-to-tail amplification across mixed traffic compositions\label{s7.2.2}}

Figure~\ref{fig:follower_level_stability}(d)--(i) extends the stability analysis from individual followers to heterogeneous platoons. The platoon-level response is evaluated by the product-based head-to-tail amplification metric \(A_{\mathrm{HTS}}(\omega;\mathcal{X}_0)\) and its maximum value~\(A_{\mathrm{HTS}}^{\max}\). For each feasible \((P,O)\) composition, repeated realizations are used to account for stochastic vehicle ordering and conditional parameter sampling. Therefore, the results reflect not only the local response of each interaction type, but also how these responses are arranged along the vehicle sequence.

The unstable frequency bands in Fig.~\ref{fig:follower_level_stability}(g)--(i) contract toward lower frequencies as speed increases, with the median upper band limit decreasing from about 0.11~rad/s at 10~m/s to about 0.011 and 0.004~rad/s at 20 and 30~m/s, respectively. This is consistent with the follower-level results in Section~\ref{s7.2.1}: amplification, when present, is mainly a low-frequency phenomenon, and the relevant frequency range shifts downward as the equilibrium speed increases. A similar frequency pattern has been reported in empirical studies. Observed traffic oscillations typically have periods of several minutes~\citep{mauch2002freeway, schonhof2007empirical}, and their wavelength and period become shorter as the speed in congested traffic decreases, while their propagation speed remains nearly constant~\citep{treiber2011evidence}. This corresponds to the broader unstable band under the 10~m/s condition and its contraction toward lower frequencies as the equilibrium speed increases.

The magnitude of head-to-tail amplification also decreases with speed. In Fig.~\ref{fig:follower_level_stability}(d)--(f), the 95th percentile of~\(A_{\mathrm{HTS}}^{\max}\), denoted as \(A_{\mathrm{HTS}}^{\mathrm{p95}}\), approaches 3.5 in the high-amplification region at 10~m/s. This indicates that follower-level gains that are individually close to one can still accumulate into substantial platoon-level amplification over a long vehicle sequence. At 20~m/s, the maximum \(A_{\mathrm{HTS}}^{\mathrm{p95}}\) decreases to about 1.4, while at 30~m/s it remains close to the neutral threshold and stays below about 1.05 over the plotted space. The decline of amplification with speed is consistent with empirical evidence on traffic instability. In fact, oscillations have been observed to arise mainly in queued, low-speed traffic~\citep{mauch2002freeway}, and perturbations generally grow in congested traffic but decay in free flow~\citep{treiber2011evidence}. Platoon experiments and freeway data further identify a critical speed of about 8--11~m/s, below which traffic flow tends to be unstable~\citep{jiang2018experimental}, and experiments with commercial ACC vehicles show that disturbance amplification is more likely at lower speeds~\citep{li2021car}.

The composition pattern is mainly governed by the joint effect of AV penetration and vehicle ordering. Higher amplification is generally concentrated in the high-\(P\), positive-\(O\) region, especially under the 10 and 20~m/s conditions. As \(P\) increases, AV-follower interactions become more frequent. When \(O\) is also positive, similar vehicle types are more likely to appear consecutively, increasing the occurrence of AV-related interaction sequences, especially AV--AV sequences. Because these repeated interactions may share overlapping low-frequency amplification tendencies, their responses can accumulate more strongly along the platoon. By contrast, more alternating arrangements mix different interaction types and can partly interrupt this accumulation.

Further experimental results are included in Appendix~\ref{app:results_stability}, where we show the same speed dependence in terms of occurrence, as well as time-domain simulations with sinusoidal leader perturbations.

\subsection{Ring-road flow evolution under periodic-boundary conditions\label{s7.3}}

\subsubsection{Fundamental diagram and critical density\label{s7.3.1}}

\begin{figure}[pos=t]
\centering
\includegraphics[width=\linewidth]{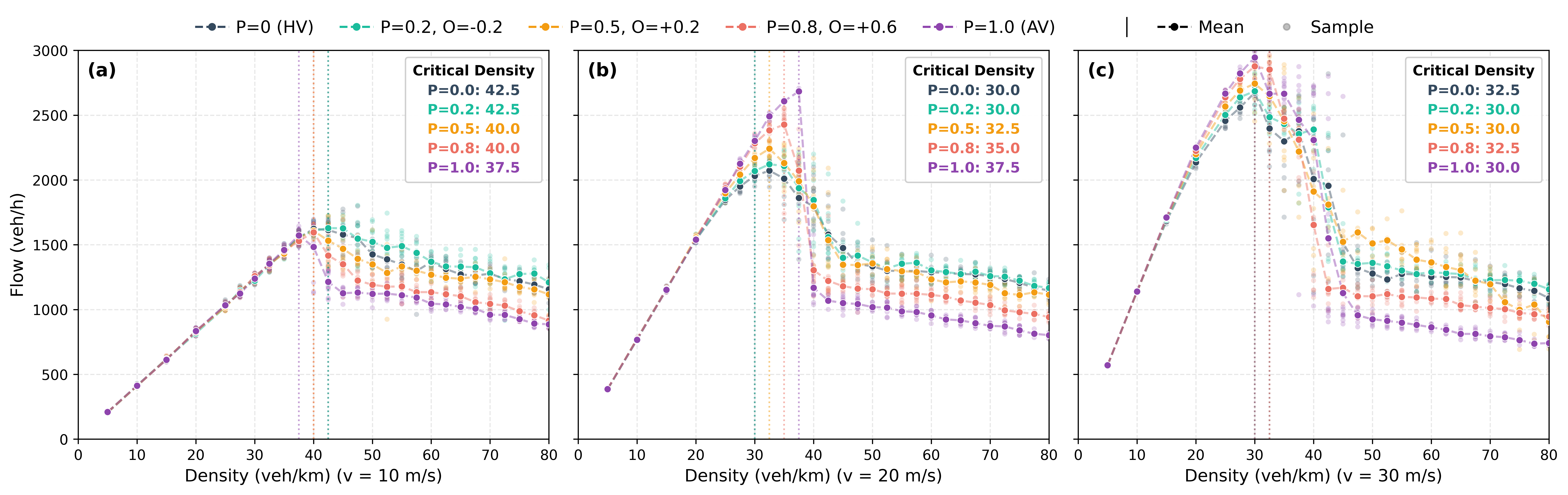}
\caption{Fundamental diagrams from ring-road simulations under periodic-boundary conditions. Panels (a)--(c) correspond to speed conditions of 10, 20, and 30~m/s, respectively. Connected markers show mean flow values across repeated realizations, scattered points show realization-level samples, and vertical dashed lines indicate the critical density associated with the capacity point.}
\label{fig:ring_fd}
\end{figure}

Figure~\ref{fig:ring_fd} presents the density--flow relationships obtained from the periodic-boundary ring-road simulations. Different from the equilibrium-flow calculation at prescribed speeds in Section~\ref{s7.1}, the flow values here are realized after density-controlled ring-road evolution and are aggregated over repeated realizations. Here, the speed condition is the speed used to generate the behavioral parameters, not the operating speed, which is determined by the density. At each density, the parameters are generated from the prescribed speed and the corresponding initial time headway and then kept fixed during the simulation. Each panel therefore shows how a traffic stream whose behavior is generated at one speed level evolves over the full density range, and its congested branch reflects the low-speed response of these parameters. The scattered points show the realization-level variability caused by stochastic vehicle ordering and conditional parameter sampling, while the connected markers summarize the corresponding condition-level mean trends. The selected curves provide representative comparisons across pure HV traffic, mixed compositions, and pure AV traffic, and the critical density over the full feasible \((P,O)\) space is provided in Appendix~\ref{app:results_ring}.

Across the three speed conditions, the fundamental diagrams show a consistent increase--peak--decline pattern. In the low-density range, the curves for different compositions remain close to each other, indicating that composition has limited influence when the density constraint is weak. As density approaches the capacity region, the curves begin to separate. After the capacity point is exceeded, the post-critical branches decline and the realization-level scatter becomes more evident. This indicates that mixed traffic composition affects not only the attainable peak flow, but also the variability and degradation of realized flow after the system enters the post-critical regime.

The speed condition changes both the peak-flow level and the shape of the post-critical branch. Under the 10~m/s condition, the peak flow remains relatively low, and increasing AV penetration does not produce a clear improvement over the pure HV case. Under the 20~m/s condition, the displayed high-\(P\), positive-\(O\) and pure-AV cases reach higher peak flows, and the separation among compositions becomes clearer around the capacity region. Under the 30~m/s condition, the ascending branch can reach a high flow level, but the post-critical decline is sharper. Thus, higher-speed traffic states can support higher potential flow, but they also become more sensitive to density increases and mean-speed reductions after the capacity point is exceeded.

These patterns are broadly consistent with the equilibrium-flow results in Section~\ref{s7.1}. The low-speed ring-road results again show that AV penetration does not necessarily increase capacity, whereas the medium- and high-speed results are consistent with the shorter equilibrium headways associated with AV-follower interactions. However, the fundamental diagrams from the ring-road simulations add a density-evolution perspective that is not captured by the prescribed-speed equilibrium-flow estimate alone. A composition that raises the peak flow may still experience a faster post-critical flow reduction once the density exceeds the capacity point.

Taken together, the fundamental diagrams from the ring-road simulations show how the two mechanisms combine. The capacity point is mainly governed by the equilibrium headways at medium and high speeds: under the 20 and 30~m/s conditions, where AV-follower interactions have shorter THWs (Section~\ref{s7.1}), compositions with higher \(P\) reach higher peak flows. The congested branch is instead governed by the headways at low speeds, because post-critical states operate at reduced speeds where AV-follower interactions have longer THWs; accordingly, the pure-AV congested branch is the lowest in all three panels of Fig.~\ref{fig:ring_fd}. The transition between the two is shaped by disturbance amplification: compositions with higher \(P\) and positive \(O\), which accumulate stronger head-to-tail amplification (Section~\ref{s7.2.2}), show a sharper post-critical flow reduction.

\subsubsection{Space--time evolution of post-critical low-speed waves\label{s7.3.2}}

\begin{figure}[pos=t]
\centering
\includegraphics[width=\linewidth]{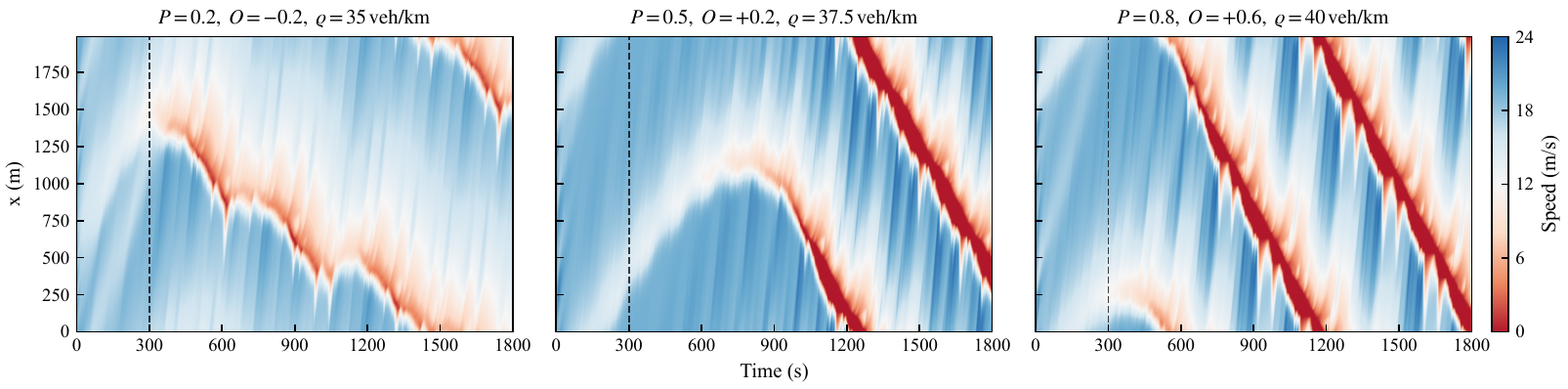}
\caption{Space--time speed evolution under periodic-boundary ring-road simulations: composition comparison at $20~\mathrm{m/s}$, using representative density conditions around $\varrho_c+5~\mathrm{veh/km}$. The dashed vertical line marks the end of the warm-up period. Colors indicate speed, with a shared color scale across panels. Each panel shows one representative realization.}
\label{fig:ring_st_evolution}
\end{figure}

Figure~\ref{fig:ring_st_evolution} complements the fundamental diagrams by showing how the post-critical low-flow transition develops in the periodic-boundary ring-road simulations. Instead of providing another aggregate density--flow metric, the space--time plots show how speed reductions are organized after the system operates at representative densities above the corresponding critical density. The figure compares three representative compositions under the 20~m/s speed condition; a comparison across speed conditions for a fixed composition is provided in Appendix~\ref{app:results_ring}. Since each panel shows one representative realization, the figure is used to illustrate dynamic organization, while the aggregate patterns are provided by Fig.~\ref{fig:ring_fd}.

Across the selected cases, localized speed reductions emerge after the warm-up period and develop into oblique low-speed bands in the space--time domain. Under periodic-boundary conditions, these low-speed regions persist and re-enter the observation domain, producing recurrent low-speed wave patterns rather than a uniform reduction in speed. This explains the post-critical flow reduction observed in Fig.~\ref{fig:ring_fd}: once the density exceeds the capacity point, part of the ring is repeatedly occupied by low-speed regions, which reduces the realized mean speed and flow.

Figure~\ref{fig:ring_st_evolution} further suggests that mixed traffic composition affects the organization of the post-critical waves. Along the selected sequence from \((P,O)=(0.2,-0.2)\) to \((0.5,0.2)\) and \((0.8,0.6)\), the low-speed bands become more coherent and continuous. The corresponding branch-based wave-speed estimates are \(-1.824\), \(-2.964\), and \(-3.065~\mathrm{m/s}\), respectively. This pattern indicates that, under the selected 20~m/s post-critical density conditions, higher \(P\) and positive \(O\) are associated with more organized upstream-moving low-speed wave structures. Because \(P\), \(O\), and the selected post-critical density condition vary together in this comparison, the result should be interpreted as representative evidence of post-critical wave organization rather than as an isolated sensitivity analysis of either \(P\) or \(O\).

\section{Discussion\label{s8}}

\subsection{Equilibrium headway and disturbance response\label{s8.1}}

The results can be interpreted through two mechanisms that transmit interaction-specific car-following behavior to the system level. The first is the equilibrium-headway mechanism. Because AV-follower interactions have longer THWs than HV-follower interactions at low speed but shorter THWs at medium and high speeds, increasing \(P\) reduces equilibrium flow at 10~m/s but raises equilibrium flow and ring-road capacity at 20 and 30~m/s, while \(O\) modifies these effects by changing whether similar interaction types are repeated or interrupted. This interpretation is consistent with mixed traffic studies that relate capacity changes to headway settings, platoon organization, and vehicle-type composition~\citep{chen2017towards, ghiasi2017mixed, sala2021capacity}. The second is the disturbance-response mechanism. Follower-level amplification is generally mild, but because platoon-level transmission depends on the product of follower-level responses, small low-frequency gains accumulate when AV-related interactions are frequent and occur consecutively, as in high-\(P\), positive-\(O\) compositions. Penetration rate alone is therefore insufficient to characterize disturbance propagation, consistent with findings that ACC or AV responses vary across situations and that oscillation growth depends on AV position in mixed platoons~\citep{yu2025dynamic, zheng2025oscillation}.

The two mechanisms combine in the fundamental diagram from the ring-road simulations and imply a trade-off between efficiency and stability. Compositions with higher \(P\) and positive \(O\) raise the capacity point at medium and high speeds through shorter headways, but once the critical density is exceeded, they show a sharper flow reduction, organized through recurrent upstream-moving low-speed waves rather than a uniform speed reduction. A composition that improves capacity can therefore still move the system closer to disturbance amplification and post-critical breakdown, consistent with ring-road and stop-and-go-wave studies emphasizing recurrent disturbance propagation under periodic traffic dynamics~\citep{stern2018dissipation, luo2024stabilizing}.

\subsection{Why reported effects of automation differ across studies\label{s8.2}}

These mechanisms also help explain why the reported effects of automation differ across mixed traffic studies. Analyses based on connected or cooperative vehicle assumptions, short intra-platoon headways, or explicitly stabilizing control strategies often report capacity or stability improvements~\citep{talebpour2016influence, ghiasi2017mixed, liu2018modeling}. Empirical studies of commercial ACC or AV behavior, by contrast, have reported conservative headways, response delays, asymmetric acceleration--deceleration behavior, and string-unstable responses~\citep{shang2021impacts, shi2021empirical, yu2025dynamic}. The present study does not assume cooperative vehicle control or platoon coordination, so its results reflect the behavior of currently observed AVs. Because observed AV behavior depends on headway settings and speed levels~\citep{li2021car, shi2021empirical, hu2023autonomous}, and AV trajectories at different speeds are often collected in different operating environments, interaction-type differences estimated from trajectory data are partly confounded with these environments. For instance, the shorter AV-follower THWs at medium and high speeds may partly reflect the highway sources of high-speed AV trajectories. Studies that draw AV behavior from different sources may therefore reach different conclusions even under similar penetration rates.

\subsection{Implications for current automated vehicles\label{s8.3}}

For currently deployed, non-cooperative AVs, the results suggest three implications for longitudinal behavior. First, for stability, the main room for improvement lies in the disturbance response rather than in the headway policy. The observed AV headways are longer at low speed, where amplification is strongest, and shorter at medium and high speeds, where amplification is weak. However, under the 20 and 30~m/s conditions, AV-follower interactions still show a larger share of amplifying samples than HV-follower interactions (Fig.~\ref{fig:stability_supplement}(b)), and under the 10~m/s condition, amplification accumulates strongly along the platoon for both follower types. Lengthening the headway further would therefore lower capacity without necessarily removing this amplification. Since capacity is set by the equilibrium headway, whereas amplification depends on how a follower responds to speed changes of its leader, improving this response can reduce amplification without a loss of capacity. The amplified disturbances are slow, with periods of about one minute or longer: the unstable band extends to about 0.11~rad/s at 10~m/s and is confined below about 0.011 and 0.004~rad/s at 20 and 30~m/s, respectively. Detecting such slow disturbances requires long observation windows and long platoons, which short experiments may not provide. The string-stability assessment of AVs should therefore include slow, periodic speed disturbances.

Second, AVs do not raise capacity under the congested, low-speed conditions where additional capacity is most needed. The capacity gains at medium and high speeds come from shorter equilibrium headways, whereas in congested and queued traffic, flow is governed by low-speed headways, at which AV followers keep longer headways than HV followers. Accordingly, the pure-AV congested branch is the lowest under all three speed conditions (Fig.~\ref{fig:ring_fd}). The low-speed headway is therefore a key AV behavioral parameter for congested traffic, and its balance against safety and comfort should be considered explicitly in AV design.

Third, clustering strengthens the existing response of AV-follower interactions. Because head-to-tail amplification is the product of follower-level responses, consecutive AV-follower interactions accumulate amplification when their gains exceed one and accumulate attenuation when their gains are below one. The capacity benefit of dedicated AV lanes or other arrangements that group AVs together, which arises mainly at medium and high speeds, therefore depends on AV-follower interactions being string stable at the speeds where such arrangements operate. Otherwise, clustering strengthens amplification, as reflected by the 95th-percentile head-to-tail amplification of up to about 3.5 at 10~m/s and about 1.4 at 20~m/s in high-\(P\), positive-\(O\) compositions (Fig.~\ref{fig:follower_level_stability}(d)--(e)).

\section{Conclusion\label{s9}}

This study developed a trajectory-informed and traffic-state-conditioned framework for parameterizing heterogeneous car-following behavior in mixed AV--HV traffic. The calibrated IDM parameters reproduced short-horizon car-following behavior and preserved interaction-specific heterogeneity, and the conditional representation generated behavioral inputs that reproduced the main trajectory-level indicator distributions, providing a behaviorally grounded input for system-level mixed traffic analysis.

The system-level experiments showed that the effect of AV penetration is conditional rather than uniformly positive and operates through two mechanisms. Through the equilibrium headway, higher AV penetration reduced equilibrium flow at low speed but increased equilibrium flow and (ring-road) capacity at medium and high speeds. Through the disturbance response, small low-frequency gains accumulated along platoons when AV-related interactions were frequent and occurred consecutively, most strongly at low speed. In the fundamental diagram from the ring-road simulations, compositions with higher penetration and positive clustering raised the capacity point at medium and high speeds but showed a sharper post-critical flow reduction through persistent upstream-moving low-speed waves.

These findings indicate that mixed traffic evolution cannot be characterized by AV penetration rate alone: the penetration rate determines how often AVs appear, the clustering intensity determines whether adjacent vehicles form same-type (AV--AV and HV--HV) or mixed (AV--HV and HV--AV) interaction types, and the speed condition determines how the differences between interaction types are expressed. Capacity gains and stability are therefore related but not equivalent outcomes, and comparing them requires behavioral inputs that are conditioned on the local traffic state rather than fixed or transferred parameters.

Several limitations remain. The study focuses on single-lane longitudinal car-following and does not include lane changing, merging, bottleneck operation, or network-level effects. The AV behavior is represented by trajectory-informed car-following observed in available data rather than by cooperative or connected control. In addition, the deployed parameters are kept fixed during each simulation after being generated from the initial traffic state. Future work can extend the framework to multi-lane and network-level settings, incorporate lane-changing and merging interactions, and examine cooperative or connected AV behavior when suitable empirical data become available. Broader empirical observations of mixed traffic would also help validate the system-level patterns identified here.

\section*{CRediT authorship contribution statement}
Hao Lyu: Conceptualization, Methodology, Software, Data curation, Formal analysis, Visualization, Writing -- original draft. Yanyong Guo: Supervision, Funding acquisition, Writing -- review and editing. Claudio Roncoli: Methodology, Supervision, Writing -- review and editing.

\section*{Acknowledgement}
This research was supported by the National Key R\&D Program of China (No. 2023YFB4302701), the National Natural Science Foundation of China (No. 52232012, 52272343, 52131203), the China Scholarship Council (No. 202506090183), Postgraduate Research \& Practice Innovation Program of Jiangsu Province (No. KYCX25\_0502) and the SEU Innovation Capability Enhancement Plan for Doctoral Students (No. CXJH\_SEU 25178).

\section*{Declaration of generative AI and AI-assisted technologies in the manuscript preparation process}
During the preparation of this work, the authors used Claude (Anthropic) to improve the language, readability, and consistency of the manuscript. After using this tool, the authors reviewed and edited the content as needed and take full responsibility for the content of the published article.

\begin{appendices}

\section{Detailed calculation of system-level evaluation metrics\label{app:system_metrics}}

This appendix details the calculations and aggregation procedures for the system-level metrics defined in Section~\ref{s5.3}.

\subsection{Equilibrium headways\label{app:metric_headways}}

Following the headway-based formulation for Markov-chain mixed traffic~\citep{ghiasi2017mixed}, the equilibrium time headway of follower \(n\) at speed \(v\) is calculated from the equilibrium gap in Eq.~\eqref{eq:ch5_equilibrium_spacing}:
\begin{equation}
h_n^{\mathrm{eq}}=
\frac{s_n^{\mathrm{eq}}+l_{\mathrm{veh}}}{v},
\label{eq:ch5_equilibrium_headway}
\end{equation}
where \(l_{\mathrm{veh}}\) denotes the vehicle length. For each interaction type \(r\), the mean equilibrium headway is
\begin{equation}
\bar{h}_{r}^{\mathrm{eq}}(v)
=
\mathbb{E}\!\left(
h_n^{\mathrm{eq}}
\mid r_n=r,\, v_n=v
\right),
\qquad r\in\mathcal{R}.
\label{eq:ch5_type_headway}
\end{equation}
These type-specific headways enter the equilibrium-flow calculation in Eq.~\eqref{eq:ch5_capacity}; finite-realization interaction shares are obtained from Eq.~\eqref{eq:ch5_interaction_share}.

\subsection{Analytical stability calculations\label{app:metric_stability}}

Following the linearization and frequency-domain treatment of car-following stability~\citep{feng_string_2019, montanino_string_2021}, each follower is linearized around the equilibrium state determined by \(s_n^{\mathrm{eq}}\), \(v\), and \(\Delta v_n=0\). The partial derivatives of the IDM acceleration function \(f=f_{\mathrm{IDM}}\) are
\begin{equation}
f_{v,n}
=
\left.\frac{\partial f}{\partial v_n}\right|_{\mathrm{eq}},
\qquad
f_{s,n}
=
\left.\frac{\partial f}{\partial s_n}\right|_{\mathrm{eq}},
\qquad
f_{\Delta v,n}
=
\left.\frac{\partial f}{\partial \Delta v_n}\right|_{\mathrm{eq}}.
\label{eq:ch5_stability_partials}
\end{equation}
Since the relative velocity is defined as \(\Delta v_n(t)=v_n(t)-v_{n-1}(t)\), the follower-level transfer function is
\begin{equation}
G_n(\lambda)
=
\frac{f_{s,n}-f_{\Delta v,n}\lambda}
{\lambda^2-\left(f_{v,n}+f_{\Delta v,n}\right)\lambda+f_{s,n}},
\qquad
\lambda=j\omega ,
\label{eq:ch5_single_vehicle_transfer}
\end{equation}
where \(j\) is the imaginary unit and \(\omega\) is the disturbance frequency. This transfer function is used to calculate the single-follower gain in Eq.~\eqref{eq:ch5_single_vehicle_gain} and the platoon head-to-tail spectrum in Eq.~\eqref{eq:ch5_hts_spectrum}.

For the frequency grid \(\Omega\) and tolerance \(\epsilon_A=0.001\), the unstable frequency set and realization-level instability indicator are
\begin{equation}
\Omega_{\mathrm{unstable}}(\mathcal{X}_0)
=
\left\{
\omega\in\Omega:
A_{\mathrm{HTS}}(\omega;\mathcal{X}_0)>1+\epsilon_A
\right\},
\qquad
\mathbb{I}_{\mathrm{unstable}}(\mathcal{X}_0)
=
\mathbb{I}\!\left(
A_{\mathrm{HTS}}^{\max}(\mathcal{X}_0)>1+\epsilon_A
\right),
\label{eq:ch5_unstable_set}
\end{equation}
where \(A_{\mathrm{HTS}}^{\max}\) is defined in Eq.~\eqref{eq:ch5_hts_max} and \(\mathbb{I}(\cdot)\) is the indicator function. For \(B\) repeated realizations under a traffic condition \(\kappa=(P,O,v)\), the unstable rate is
\begin{equation}
R_{\mathrm{unstable}}(\kappa)
=
\frac{1}{B}
\sum_{b=1}^{B}
\mathbb{I}_{\mathrm{unstable}}
\!\left(\mathcal{X}_0^{(b)}\right).
\label{eq:ch5_unstable_rate}
\end{equation}

\subsection{Sinusoidal response amplitudes\label{app:metric_sinusoidal}}

Open-boundary sinusoidal perturbation simulations are used as a time-domain verification of the analytical stability metrics. A sinusoidal speed disturbance is imposed on the leading vehicle, while the following vehicles evolve according to the IDM with the deployed behavioral parameters. During the forcing interval, the leader speed is specified as
\begin{equation}
v_1(t)
=
v+\alpha_v v
\sin\!\left(\omega(t-t_{\mathrm{s}})\right),
\qquad
t_{\mathrm{s}}\leq t\leq t_{\mathrm{s}}+T_{\mathrm{f}},
\label{eq:ch5_sine_input}
\end{equation}
where \(\alpha_v\) is the relative disturbance amplitude, \(t_{\mathrm{s}}\) is the perturbation starting time, and \(T_{\mathrm{f}}\) is the forcing duration. Outside this interval, the leader speed is set to \(v\).

The response to the sinusoidal leader disturbance in Eq.~\eqref{eq:ch5_sine_input} is evaluated at the imposed frequency, consistent with frequency-domain string-stability analysis~\citep{naus2010string, feng_string_2019}. Over the response evaluation window \(\mathcal{T}_{\mathrm{resp}}\), the amplitude of vehicle \(n\) is computed as
\begin{equation}
\mathcal{A}_n(\omega)
=
2
\sqrt{
\left[
\frac{1}{|\mathcal{T}_{\mathrm{resp}}|}
\int_{\mathcal{T}_{\mathrm{resp}}}
\tilde v_n(t)
\cos\!\left(\omega(t-t_{\mathrm{s}})\right)
\,dt
\right]^2
+
\left[
\frac{1}{|\mathcal{T}_{\mathrm{resp}}|}
\int_{\mathcal{T}_{\mathrm{resp}}}
\tilde v_n(t)
\sin\!\left(\omega(t-t_{\mathrm{s}})\right)
\,dt
\right]^2
},
\label{eq:ch5_lockin_amplitude}
\end{equation}
where \(\tilde v_n(t)=v_n(t)-\bar v_n\), \(\bar v_n\) is the mean speed over the response window, and \(|\mathcal{T}_{\mathrm{resp}}|\) is the duration of that window.

The simulation-based tail gain is then defined as
\begin{equation}
G_{\mathrm{sim}}(\omega;\mathcal{X}_0)
=
\frac{\mathcal{A}_{N}(\omega)}
{\mathcal{A}_{1}(\omega)} .
\label{eq:ch5_sim_gain}
\end{equation}
A value of \(G_{\mathrm{sim}}>1\) indicates that the imposed disturbance is amplified from the leader to the tail vehicle in the open-boundary simulation. The comparison between \(A_{\mathrm{HTS}}(\omega;\mathcal{X}_0)\) at the imposed frequency and \(G_{\mathrm{sim}}(\omega;\mathcal{X}_0)\) is used to assess whether the analytical amplification tendency is reflected in dynamic vehicle trajectories.

\subsection{Ring-road aggregation and wave-speed estimation\label{app:metric_ring}}

The ring-road analyses use the density--flow relationship, the capacity point, and space--time speed evolution introduced in Section~\ref{s5.3.4}~\citep{zhou2020modeling, yao2022fundamental, sun2024fundamental}. For each density \(\varrho\) and condition \(\kappa=(P,O,v)\), the mean speed after the warm-up period is
\begin{equation}
\bar{v}^{(b)}(\varrho;\kappa)
=
\frac{1}{N|\mathcal{T}_{\mathrm{stat}}|}
\sum_{n=1}^{N}
\sum_{t\in\mathcal{T}_{\mathrm{stat}}}
v_{n}^{(b)}(t),
\label{eq:ch5_ring_mean_speed}
\end{equation}
where \(|\mathcal{T}_{\mathrm{stat}}|\) denotes the number of sampled time steps in the statistical evaluation window. The realization-level flow is calculated from Eq.~\eqref{eq:ch5_fd_flow}, and its mean over \(B\) repeated realizations is
\begin{equation}
\bar{Q}(\varrho;\kappa)
=
\frac{1}{B}
\sum_{b=1}^{B}
Q^{(b)}(\varrho;\kappa).
\label{eq:ch5_fd_mean_flow}
\end{equation}
This mean flow is used to identify the capacity point and critical density in Eq.~\eqref{eq:ch5_critical_density}.

For an identifiable mature low-speed branch in the space--time plane, the propagation speed is estimated using
\begin{equation}
x_{\mathrm{branch}}(t)
\approx
c_{\mathrm{wave}}t+x_{0},
\label{eq:ch5_wave_speed_fit}
\end{equation}
where \(x_0\) is the fitted intercept and \(c_{\mathrm{wave}}\) is the branch slope. The longitudinal coordinate increases in the vehicle moving direction, so a negative \(c_{\mathrm{wave}}\) indicates upstream propagation. The estimate is used as an auxiliary diagnostic of organized low-speed bands.

\section{Supplementary validation of behavioral parameterization\label{app:param_validation}}

\renewcommand{\thefigure}{B\arabic{figure}}
\setcounter{figure}{0}
\renewcommand{\thetable}{B\arabic{table}}
\setcounter{table}{0}

This appendix provides the numerical settings and supplementary results for the validation of behavioral parameterization in Section~\ref{s6}.

\subsection{Numerical settings\label{app:validation_settings}}

Table~\ref{tab:validation_settings} lists the IDM parameter bounds and acceleration interval, together with the main model and training settings of the calibration module and the conditional representation.

\begin{table}[pos=tbhp]
\centering
\caption{Main numerical settings for behavioral parameterization validation.}
\label{tab:validation_settings}
\small
\begin{tabular*}{\tblwidth}{@{\extracolsep{\fill}}llll@{}}
\toprule
Item & Setting & Item & Setting \\
\midrule
Parameter lower bound
& \multicolumn{3}{l}{\(\boldsymbol{\theta}_{\mathrm{L}}=(0.1,0.1,0.1,0.1,2.0,2.0)\)} \\

Parameter upper bound
& \multicolumn{3}{l}{\(\boldsymbol{\theta}_{\mathrm{U}}=(10.0,5.0,4.0,5.0,50.0,6.0)\)} \\

IDM acceleration interval
& \([-6.0,a_{\max}]\) m/s\(^2\)
& Top-\(k\) expert routing
& 2 \\

Calibration hidden dimension
& 64
& Mamba layers
& 2 \\

Attention heads
& 4
& Calibration batch size
& 256 \\

Calibration learning rate
& \(5\times10^{-4}\)
& Calibration training epochs
& 20 \\

Adaptive physical penalty coefficient
& \([0.05,0.20]\)
& Conditional latent dimension
& 32 \\

Conditional hidden dimension
& 256
& Conditional residual blocks
& 3 \\

Conditional batch size
& 4096
& Conditional learning rate
& \(3\times10^{-3}\) \\

Conditional training epochs
& 20
& KL coefficient
& Annealed to 0.05 \\
\bottomrule
\end{tabular*}
\end{table}

\subsection{Calibration accuracy by interaction type\label{app:calibration_accuracy}}

\begin{figure}[pos=tbhp]
\centering
\includegraphics[width=\linewidth]{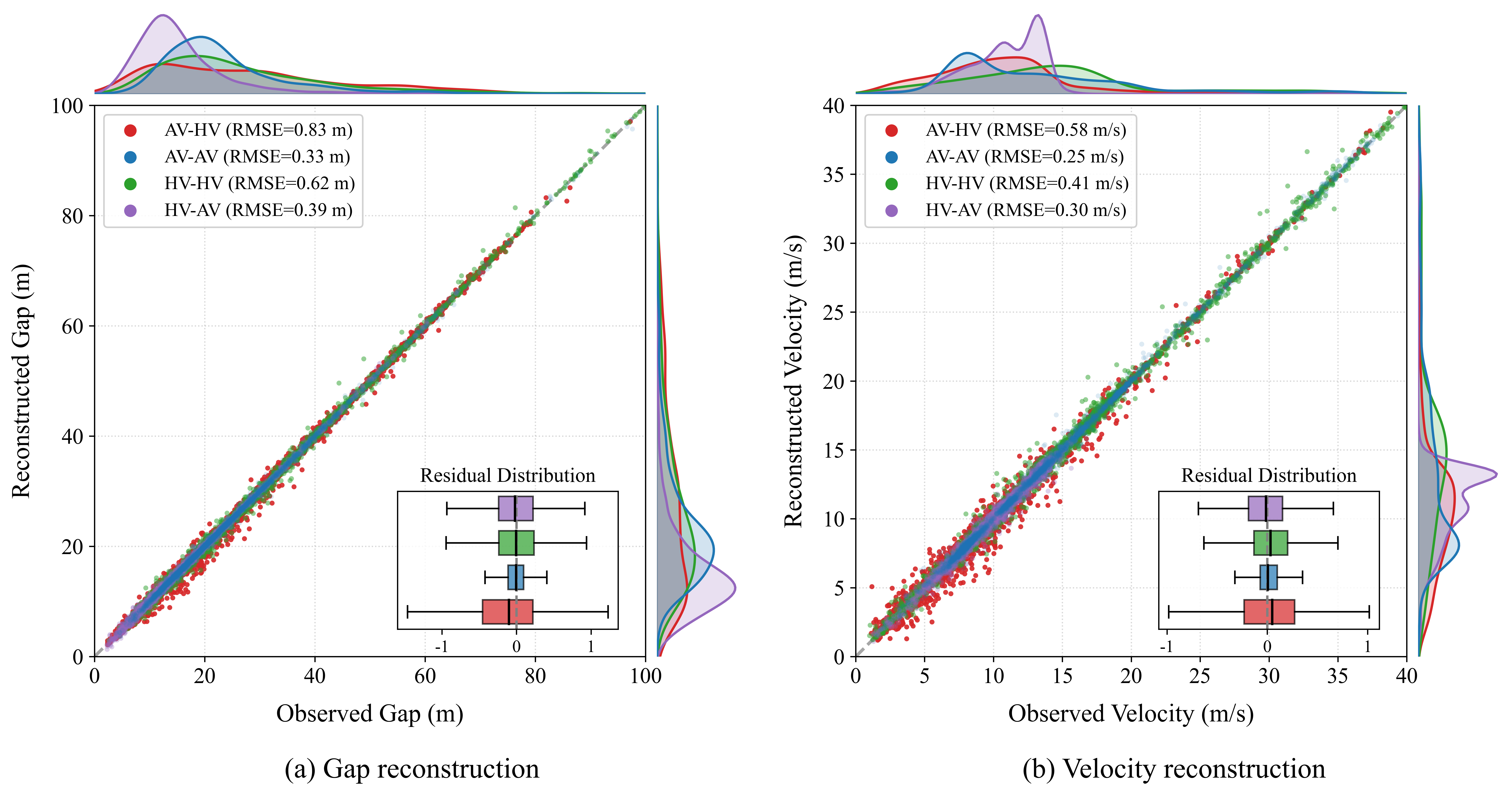}
\caption{Observed and reconstructed car-following states at the 2~s rollout endpoint using the calibrated IDM parameters: (a) gap and (b) velocity. Each interaction type contains 3,000 samples. Marginal distributions and residual distributions are shown by interaction type.}
\label{fig:calibration_scatter}
\end{figure}

Figure~\ref{fig:calibration_scatter} compares the observed and reconstructed gap and velocity values at the 2~s rollout endpoint on a balanced test subset. The residual distributions are centered close to zero for all four interaction types, and interactions with HV leaders, especially AV--HV and HV--HV, show relatively larger dispersion.

\subsection{Dataset-level distribution of comfortable deceleration\label{app:b_distribution_by_dataset}}

\begin{figure}[pos=tbhp]
\centering
\includegraphics[width=\linewidth]{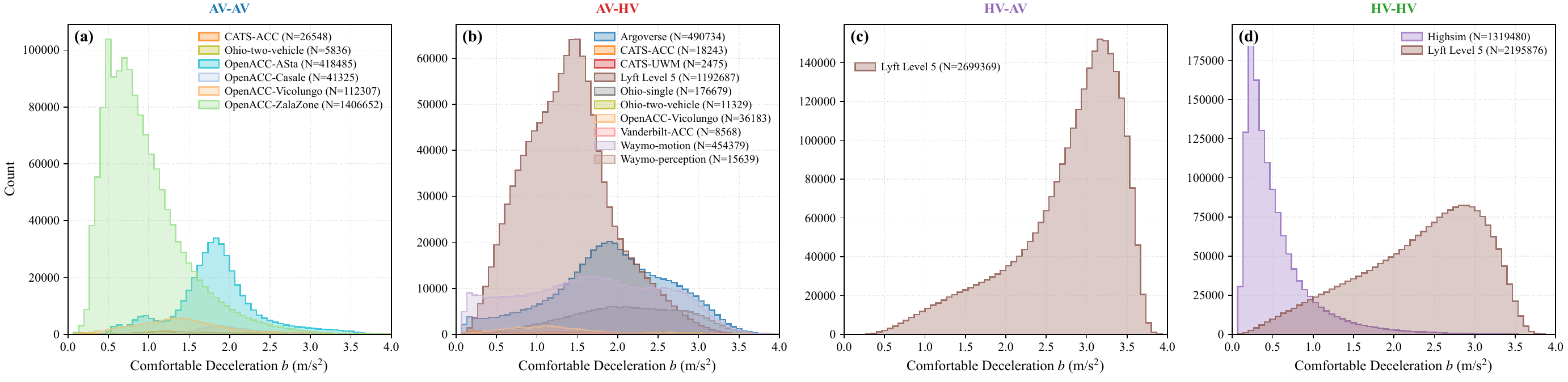}
\caption{Dataset-level distributions of the calibrated comfortable deceleration parameter $b$ by interaction type. Panels (a)--(d) show AV--AV, AV--HV, HV--AV, and HV--HV interactions, respectively. Histograms show counts in 60 equal-width bins over $0$--$4~\mathrm{m/s^2}$; $N$ denotes the number of calibrated samples from each source. The source-level differences between HIGH-SIM and Lyft Level 5 show that the calibrated $b$ values reflect operating-environment characteristics in addition to interaction-type heterogeneity.}
\label{fig:b_distribution_by_dataset}
\end{figure}

Because the HV--HV distribution in Fig.~\ref{fig:calibrated_param_kde} shows a visible bimodal shape for the calibrated comfortable deceleration parameter \(b\), Fig.~\ref{fig:b_distribution_by_dataset} further decomposes the \(b\) distribution by data source. For HV--HV interactions, the calibrated samples are mainly drawn from HIGH-SIM~\citep{shi2021video}, which provides highway trajectory data, and Lyft Level 5~\citep{mandal2020motion}, which was collected in urban and suburban environments. The calibrated \(b\) values from HIGH-SIM are concentrated in a lower range, while those from Lyft Level 5 are concentrated in a higher range. This pattern is consistent with the smoother speed adjustments in highway car following and the more frequent local speed changes in urban and suburban traffic.

For the other interaction types, source-level differences are generally smaller than the distributional differences across interaction types, and AV-follower interactions still tend to have lower \(b\) values than HV-follower interactions. Overall, the calibrated parameters reflect operating-environment characteristics in addition to reproducing trajectory data.

\subsection{Reconstructed and generated parameters\label{app:generated_parameters}}

\begin{table}[pos=tbhp]
\centering
\caption{Distributional consistency between original and reconstructed IDM parameter samples on the test set.}
\label{tab:reconstructed_param_distribution}
\footnotesize
\renewcommand{\arraystretch}{1.15}
\setlength{\tabcolsep}{3.2pt}
\begin{threeparttable}
\resizebox{\linewidth}{!}{
\begin{tabular}{llcccccccc}
\toprule
Type & Data & Joint KL & Joint JS 
& \(s_0\) & \(T\) & \(b\) & \(a_{\max}\) & \(v_0\) & \(\delta\) \\
\midrule
AV--HV 
& Orig. 
& -- & -- 
& \(2.407\pm1.455\) & \(1.987\pm0.927\) & \(1.614\pm0.727\) 
& \(0.582\pm0.400\) & \(23.746\pm6.996\) & \(3.283\pm0.515\) \\
& Recon. 
& 0.0339 & 0.0161 
& \(2.409\pm1.459\) & \(1.984\pm0.927\) & \(1.616\pm0.728\) 
& \(0.583\pm0.398\) & \(23.865\pm7.005\) & \(3.277\pm0.515\) \\
\midrule
AV--AV 
& Orig. 
& -- & -- 
& \(4.913\pm1.498\) & \(1.426\pm0.761\) & \(1.200\pm0.661\) 
& \(0.384\pm0.308\) & \(29.071\pm6.314\) & \(3.688\pm0.636\) \\
& Recon. 
& 0.0151 & 0.0059 
& \(4.916\pm1.499\) & \(1.423\pm0.761\) & \(1.199\pm0.661\) 
& \(0.386\pm0.308\) & \(29.145\pm6.368\) & \(3.683\pm0.636\) \\
\midrule
HV--HV 
& Orig. 
& -- & -- 
& \(2.425\pm1.765\) & \(1.350\pm0.896\) & \(1.652\pm1.082\) 
& \(0.816\pm0.675\) & \(21.992\pm9.379\) & \(2.967\pm0.681\) \\
& Recon. 
& 0.0118 & 0.0047 
& \(2.429\pm1.775\) & \(1.348\pm0.899\) & \(1.652\pm1.082\) 
& \(0.817\pm0.675\) & \(22.068\pm9.416\) & \(2.961\pm0.681\) \\
\midrule
HV--AV 
& Orig. 
& -- & -- 
& \(0.388\pm0.368\) & \(1.010\pm0.614\) & \(2.759\pm0.673\) 
& \(1.432\pm1.107\) & \(13.559\pm2.843\) & \(2.549\pm0.373\) \\
& Recon. 
& 0.0131 & 0.0063 
& \(0.388\pm0.366\) & \(1.007\pm0.615\) & \(2.759\pm0.674\) 
& \(1.435\pm1.107\) & \(13.619\pm2.831\) & \(2.543\pm0.375\) \\
\bottomrule
\end{tabular}
}
\vspace{5pt}
\begin{minipage}{\linewidth}
\footnotesize
\textit{Note}: Orig. and Recon. denote original and reconstructed parameter samples, respectively. IDM parameters are reported as mean \(\pm\) standard deviation. Joint KL and JS divergences are computed from the six-dimensional parameter distribution for each interaction type.
\end{minipage}
\end{threeparttable}
\end{table}

Table~\ref{tab:reconstructed_param_distribution} evaluates the traffic-state-conditioned conditional parameter representation in reconstruction mode on the held-out test set. Across the four follower--leader interaction types, the reconstructed means and standard deviations remain close to the calibrated reference values, and the joint KL and JS divergences of the six-dimensional parameter distributions are small. AV--HV shows the largest divergence, but its marginal statistics are still well preserved.

\begin{table}[pos=tbhp]
\centering
\caption{Closed-loop rollout accuracy using calibrated and generated IDM parameters by interaction type.}
\label{tab:generated_rollout_rmse}
\fontsize{8.5}{10.5}\selectfont
\renewcommand{\arraystretch}{1.15}
\setlength{\tabcolsep}{3.4pt}
\begin{threeparttable}
\begin{tabular*}{\linewidth}{@{\extracolsep{\fill}}ll|ccccc|ccccc}
\toprule
\multicolumn{2}{l|}{\multirow{2}{*}{\textbf{Interaction type and parameter source}}}
& \multicolumn{5}{c|}{\textbf{Velocity RMSE (m/s)}}
& \multicolumn{5}{c}{\textbf{Gap RMSE (m)}} \\
\cmidrule(lr){3-7} \cmidrule(lr){8-12}
Type & Source
& 0.5s & 1.0s & 1.5s & 2.0s & Avg.
& 0.5s & 1.0s & 1.5s & 2.0s & Avg. \\
\midrule
AV--HV
& Calibrated reference
& 0.409 & 0.445 & 0.501 & 0.573 & 0.482
& 0.284 & 0.470 & 0.652 & 0.848 & 0.563 \\
& Generated
& 0.327 & 0.531 & 0.699 & 0.843 & 0.600
& 0.404 & 0.768 & 1.132 & 1.509 & 0.953 \\
\midrule
AV--AV
& Calibrated reference
& 0.070 & 0.114 & 0.165 & 0.226 & 0.144
& 0.099 & 0.172 & 0.243 & 0.326 & 0.210 \\
& Generated
& 0.199 & 0.342 & 0.458 & 0.553 & 0.388
& 0.129 & 0.289 & 0.496 & 0.745 & 0.415 \\
\midrule
HV--HV
& Calibrated reference
& 0.225 & 0.279 & 0.342 & 0.424 & 0.317
& 0.235 & 0.370 & 0.488 & 0.629 & 0.431 \\
& Generated
& 0.274 & 0.459 & 0.605 & 0.723 & 0.515
& 0.297 & 0.586 & 0.909 & 1.271 & 0.766 \\
\midrule
HV--AV
& Calibrated reference
& 0.184 & 0.197 & 0.219 & 0.303 & 0.226
& 0.137 & 0.195 & 0.264 & 0.405 & 0.250 \\
& Generated
& 0.399 & 0.599 & 0.733 & 0.824 & 0.639
& 0.203 & 0.479 & 0.829 & 1.230 & 0.685 \\
\midrule
Overall
& Calibrated reference
& 0.266 & 0.300 & 0.348 & 0.419 & 0.333
& 0.214 & 0.343 & 0.466 & 0.613 & 0.409 \\
& Generated
& 0.311 & 0.497 & 0.640 & 0.753 & 0.550
& 0.287 & 0.574 & 0.895 & 1.250 & 0.751 \\
\bottomrule
\end{tabular*}
\vspace{5pt}
\begin{minipage}{\linewidth}
\footnotesize
\textit{Note}: The calibrated reference rollout represents the reconstruction performance obtained using directly calibrated IDM parameters, whereas the generated parameter rollout uses IDM parameters produced by the learned conditional representation. The overall calibrated-reference results are consistent with Table~\ref{tab:calibration_rmse}.
\end{minipage}
\end{threeparttable}
\end{table}

Table~\ref{tab:generated_rollout_rmse} reports the rollout errors by interaction type and rollout horizon. For the overall sample, the difference between generated and calibrated parameters is small at 0.5~s and widens as the horizon extends, especially for gap RMSE. Across interaction types, HV--AV shows the largest increase relative to the calibrated reference.

\section{Supplementary system-level results\label{app:system_results}}

\renewcommand{\thefigure}{C\arabic{figure}}
\setcounter{figure}{0}

This appendix provides supplementary results for the stability analysis in Section~\ref{s7.2} and the ring-road analysis in Section~\ref{s7.3}.

\subsection{Analytical stability and sinusoidal perturbation responses\label{app:results_stability}}

\begin{figure}[pos=tbhp]
\centering
\includegraphics[width=\linewidth]{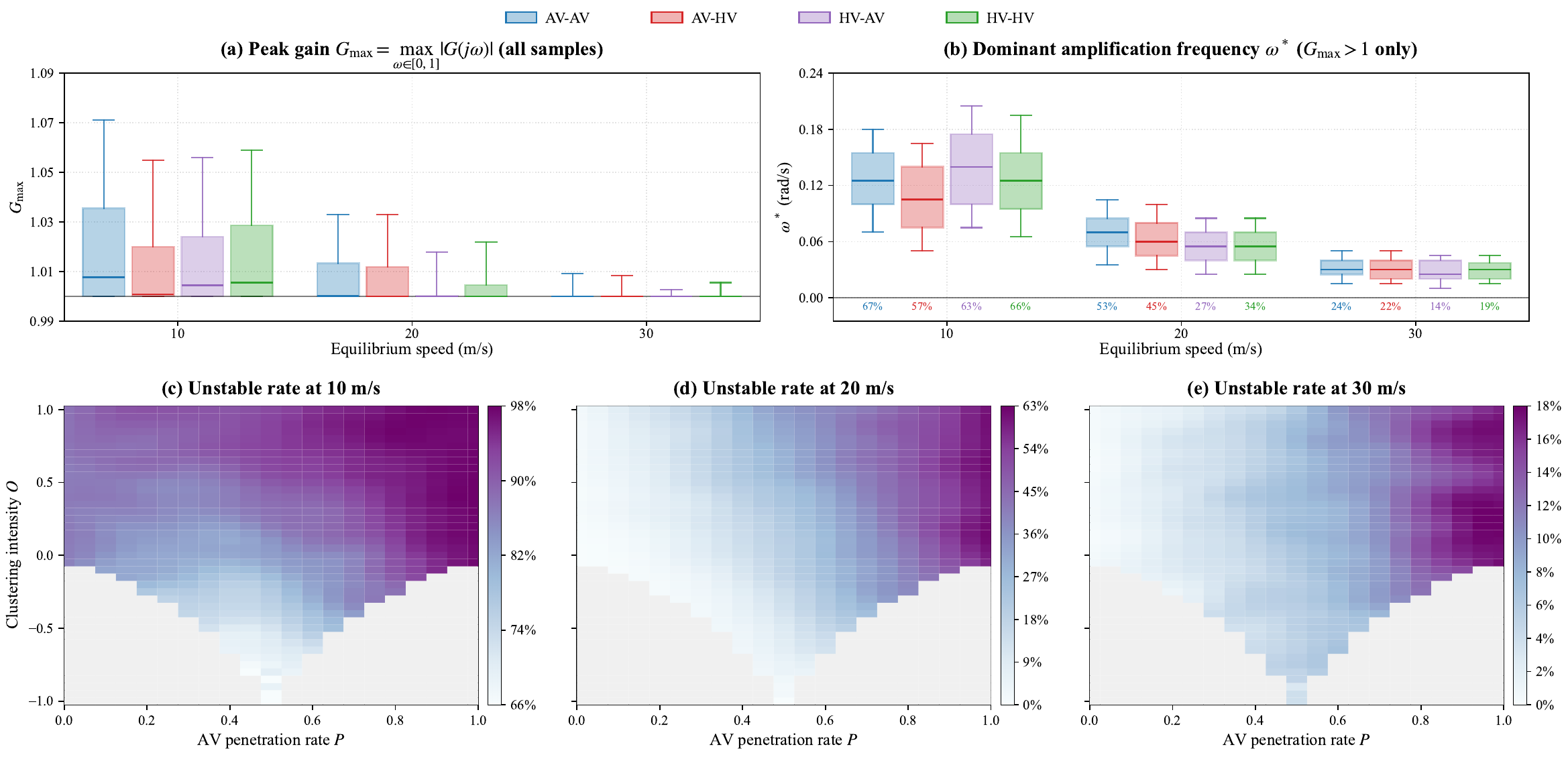}
\caption{Follower-level amplification metrics and platoon-level unstable-realization rates. Panel (a) shows the peak follower-level gain $G_{\max}$ over $\omega\in[0,1]~\mathrm{rad/s}$ for all samples, and panel (b) shows the dominant amplification frequency $\omega^*$ for samples with $G_{\max}>1$, with percentages giving the share of these samples. Boxes show medians and interquartile ranges, with whiskers at the 10th and 90th percentiles; within each speed group, interaction types are ordered AV--AV, AV--HV, HV--AV, and HV--HV, with the follower type listed first. Panels (c)--(e) show the unstable-realization rates at 10, 20, and $30~\mathrm{m/s}$, with the same realization settings and instability threshold as Fig.~\ref{fig:follower_level_stability}; gray regions indicate infeasible $(P,O)$ combinations, and color scales are set separately for each speed.}
\label{fig:stability_supplement}
\end{figure}

Figure~\ref{fig:stability_supplement}(a)--(b) summarizes the follower-level peak gain \(G_{\max}\) and the dominant amplification frequency \(\omega^*\). The peak gains remain close to one in all conditions; even under the 10~m/s condition, the 90th percentile stays below about 1.07. The share of samples with \(G_{\max}>1\) decreases with speed, from about 57--67\% at 10~m/s to 27--53\% at 20~m/s and 14--24\% at 30~m/s, and AV-follower interactions show higher shares than HV-follower interactions under the 20 and 30~m/s conditions. For amplifying samples, the dominant frequency also shifts downward with speed, with medians of roughly 0.10--0.14~rad/s at 10~m/s, 0.05--0.07~rad/s at 20~m/s, and 0.02--0.03~rad/s at 30~m/s.

The unstable-rate maps in Fig.~\ref{fig:stability_supplement}(c)--(e) provide a complementary occurrence-based view. Since the threshold \(A_{\mathrm{HTS}}^{\max}>1.001\) is used to avoid classifying numerically neutral cases as unstable, the unstable rate should be interpreted as the frequency of head-to-tail amplification occurrence rather than as its severity. The high-\(P\), positive-\(O\) tendency remains visible, but its magnitude weakens as the equilibrium speed increases.

\begin{figure}[pos=tbhp]
\centering
\includegraphics[width=\linewidth]{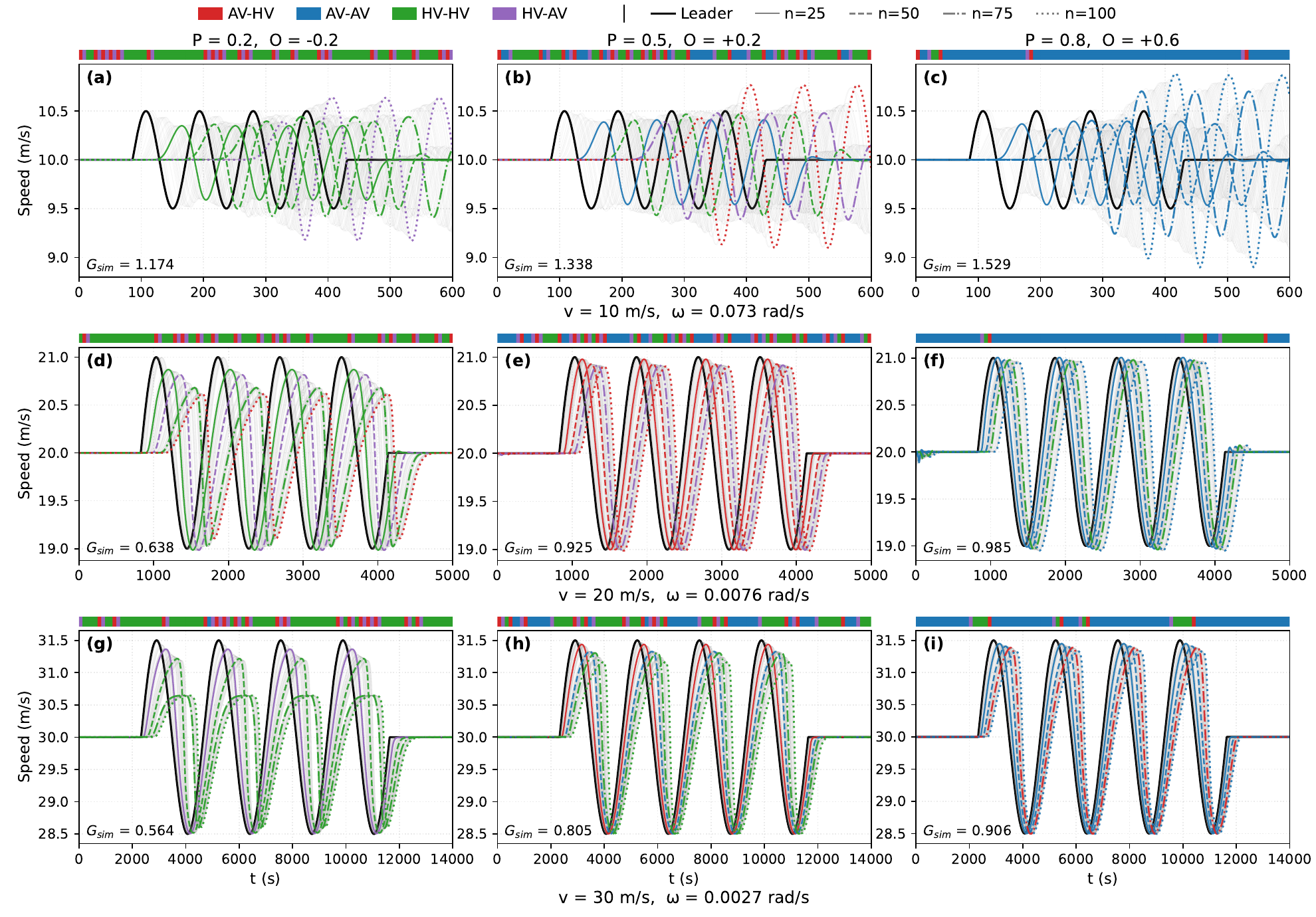}
\caption{Open-boundary speed responses under sinusoidal leader perturbations for representative mixed traffic compositions. The leader speed is perturbed with amplitude \(0.05v\). Colored bars above each panel show the realized follower--leader interaction-type ordering of the simulated sample. Highlighted trajectories correspond to the leader and followers \(n=25\), \(50\), \(75\), and \(100\). \(G_{\mathrm{sim}}\) denotes the ratio of the tail-vehicle speed amplitude to the leader speed amplitude.}
\label{fig:open_sine_response}
\end{figure}

Figure~\ref{fig:open_sine_response} verifies the analytical head-to-tail amplification results in Section~\ref{s7.2.2} through time-domain simulations of the nonlinear IDM with the deployed parameters. For each speed condition, the forcing frequency of the leader disturbance is set to the median dominant amplification frequency of the unstable platoon realizations in Fig.~\ref{fig:follower_level_stability}(g)--(i). The same frequency is used for all three representative compositions within a given speed condition. Therefore, the differences shown in each row mainly reflect mixed traffic composition and realized interaction-type ordering, rather than changes in the imposed disturbance frequency.

The simulated responses are consistent with the analytical stability patterns. Under the 10~m/s condition, the disturbance is amplified from the leader to the tail in all selected compositions, and the amplification becomes stronger as the selected sequence moves from low \(P\) with negative \(O\) to high \(P\) with positive \(O\). Under the 20~m/s condition, the tail response remains below the neutral amplification threshold, but attenuation weakens markedly in the high-\(P\), positive-\(O\) case. The trajectories also show that disturbance transmission is not spatially uniform: local amplification and recovery may occur within the platoon even when the final tail-to-leader gain remains below one. Under the 30~m/s condition, all selected cases attenuate the imposed disturbance, although the same tendency toward weaker attenuation under high \(P\) and positive \(O\) remains visible.

\subsection{Critical density and space--time evolution\label{app:results_ring}}

\begin{figure}[pos=tbhp]
\centering
\includegraphics[width=\linewidth]{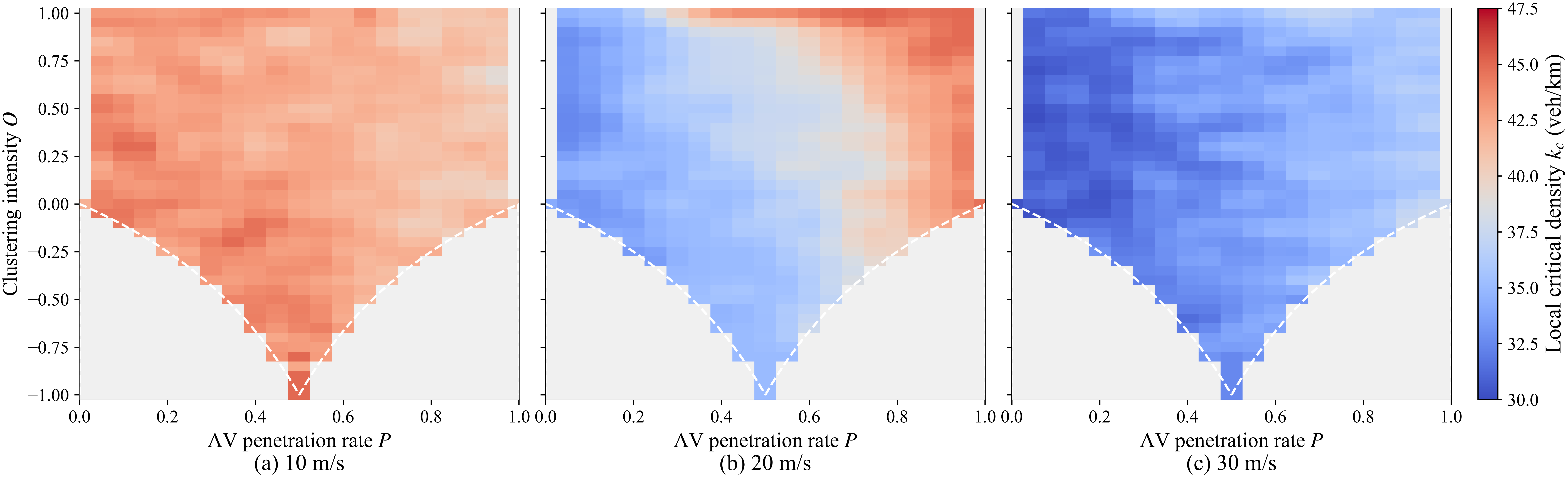}
\caption{Critical density across mixed traffic compositions under periodic-boundary ring-road simulations. Panels (a)--(c) correspond to speed conditions of 10, 20, and 30~m/s, respectively. Gray regions indicate infeasible \((P,O)\) combinations, and the dashed line marks the feasible boundary.}
\label{fig:ring_kc}
\end{figure}

Figure~\ref{fig:ring_kc} maps the critical density \(\varrho_c\) across the feasible \((P,O)\) space under periodic-boundary ring-road simulations. Here, \(\varrho_c\) is identified as the density associated with the capacity point in the aggregated density--flow relationship. Since \(\varrho_c\) describes the density location of the capacity point, it should be interpreted together with the corresponding flow levels shown in Fig.~\ref{fig:ring_fd}, rather than as a standalone capacity magnitude. As the prescribed speed condition increases from 10 to 30~m/s, the overall level of \(\varrho_c\) becomes lower. From a spacing perspective, a given time headway corresponds to a larger spatial gap at higher speed, so compact high-speed states reach the capacity point at lower densities.

The effect of \(P\) is also speed dependent. Under the 10~m/s condition, \(\varrho_c\) generally decreases as \(P\) increases. Under the 20 and 30~m/s conditions, the direction is reversed, and higher \(P\) tends to support a higher critical density, more clearly at 20~m/s than at 30~m/s. This pattern is consistent with the equilibrium-flow results in Section~\ref{s7.1}.

The influence of \(O\) is weaker and less uniform than that of \(P\). Under the 10~m/s condition, the effect of \(O\) does not follow a simple monotonic pattern. Under the 20 and 30~m/s conditions, positive \(O\) more often increases \(\varrho_c\), especially at intermediate or high penetration rates where vehicle ordering has a stronger influence on the realized follower--leader interaction sequence.

\begin{figure}[pos=tbhp]
\centering
\includegraphics[width=\linewidth]{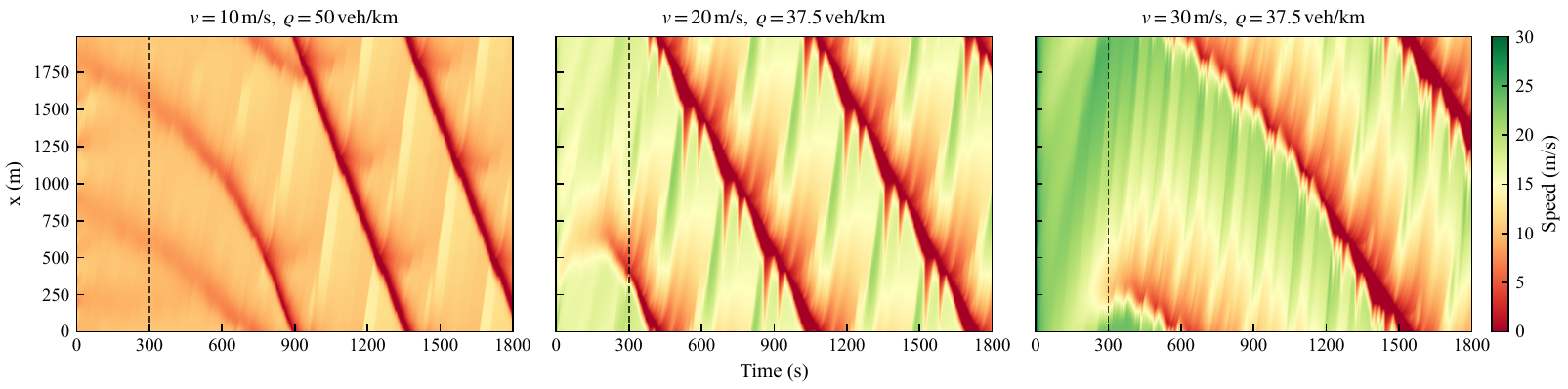}
\caption{Space--time speed evolution under periodic-boundary ring-road simulations: speed-condition comparison for $(P,O)=(0.2,-0.2)$, using representative density conditions around $\varrho_c+7.5~\mathrm{veh/km}$. The dashed vertical line marks the end of the warm-up period. Colors indicate speed, with a shared color scale across panels. Each panel shows one representative realization.}
\label{fig:ring_st_speed_effect}
\end{figure}

The speed-condition comparison in Fig.~\ref{fig:ring_st_speed_effect} shows that post-critical low-speed waves appear under all three speed conditions. The 10~m/s case has a lower background speed and a more moderate speed contrast, whereas the 20 and 30~m/s cases show sharper separation between high-speed and low-speed regions after the transition. For the mature low-speed branches identifiable in these representative realizations, the branch-based wave-speed estimates are \(-4.256\), \(-3.057\), and \(-2.531~\mathrm{m/s}\) for the 10, 20, and 30~m/s conditions, respectively. These estimates confirm the upstream-moving nature of the organized low-speed waves, while also showing that stronger post-critical speed contrast at higher speed does not necessarily imply a larger absolute upstream wave speed in the selected realizations.

\clearpage
\end{appendices}

\bibliographystyle{cas-model2-names}

\bibliography{main}

\end{sloppypar}
\end{document}